%% file: D0jets_pp.tex
\documentclass[ALICE,manyauthors]{cernphprep}

\usepackage[main=american,greek]{babel}
\usepackage[T1]{fontenc}
\usepackage[utf8]{inputenc}

\usepackage[comma,square,numbers,sort&compress]{natbib}

\usepackage{hyperref}
\usepackage{subcaption}
\usepackage{mathtools}
\usepackage{amssymb,amsmath}
\usepackage{mathrsfs}
\usepackage{lineno}
\usepackage[dvipsnames]{xcolor}
\usepackage{newtxtext,newtxmath}

\makeatletter
\g@addto@macro\bfseries{\boldmath}
\makeatother

\input{commands.tex}

\begin{document}%

\begin{titlepage}
\PHyear{2019}
\PHnumber{081}      
\PHdate{25 April}  
%

\title{Measurement of the production of charm jets tagged with \Dzero\ mesons in \pp\ collisions at $\s=7$~TeV}
\ShortTitle{Production of charm jets tagged with \Dzero\ mesons}   

\Collaboration{ALICE Collaboration\thanks{See Appendix~\ref{app:collab} for the list of collaboration members}}
\ShortAuthor{ALICE Collaboration} 

\begin{abstract}
The production of charm jets in proton--proton collisions at a center-of-mass energy
of $\sqrt{s}=7$~TeV was measured with the ALICE detector at the CERN Large Hadron Collider.
The measurement is based on a data sample corresponding to a total integrated luminosity 
of $6.23$ ${\rm nb}^{-1}$, collected using a minimum-bias trigger.
Charm jets are identified by the presence of a D$^0$ meson among their constituents.
The D$^0$ mesons are reconstructed from their hadronic decay D$^0\rightarrow$K$^{-}\partgr{p}^{+}$.
The \mbox{D$^0$-meson} tagged jets are reconstructed using tracks of charged particles (track-based jets)
with the anti-$k_{\mathrm{T}}$ algorithm in the jet transverse momentum range $5<p_{\mathrm{T,jet}}^{\mathrm{ch}}<30$~${\rm GeV/}c$
and pseudorapidity $|\eta_{\rm jet}|<0.5$.
The fraction of charged jets containing a \mbox{D$^0$-meson}
increases with $p_{\mathrm{T,jet}}^{\mathrm{ch}}$ from $0.042 \pm 0.004\, \mathrm{(stat)} \pm 0.006\, \mathrm{(syst)}$ to $0.080 \pm 0.009\,
\mathrm{(stat)} \pm 0.008\, \mathrm{(syst)}$.
The distribution of D$^0$-meson tagged jets as a function of the jet 
momentum fraction carried by the D$^0$ meson in the direction of the jet axis ($z_{||}^{\mathrm{ch}}$)
is reported for two ranges of jet transverse momenta, $5<p_{\mathrm{T,jet}}^{\mathrm{ch}}<15$~${\rm GeV/}c$
and $15<p_{\mathrm{T,jet}}^{\mathrm{ch}}<30$~${\rm GeV/}c$ in the intervals $0.2<z_{||}^{\mathrm{ch}}<1.0$ and $0.4<z_{||}^{\mathrm{ch}}<1.0$, respectively.
The data are compared with results from Monte Carlo event generators (PYTHIA~6, PYTHIA~8 and Herwig~7) and with a
\mbox{Next-to-Leading-Order} perturbative Quantum Chromodynamics calculation, obtained with the POWHEG method and interfaced with PYTHIA~6
for the generation of the parton shower, fragmentation, hadronisation and underlying event.
\end{abstract}
\end{titlepage}
\setcounter{page}{2}

%
%
\input{content}               
%
%

\newenvironment{acknowledgement}{\relax}{\relax}
\begin{acknowledgement}
\section*{Acknowledgements}
\input{fa_2019-04-09.tex}    
\end{acknowledgement}

\bibliographystyle{utphys}\bibliography{biblio}

\newpage
\appendix
\section{The ALICE Collaboration}
\label{app:collab}
\input{2019-04-09-Alice_Authorlist_2019-Apr-09.tex}  
\end{document}

%% file: commands.tex
\newcommand{\kap}{K$^{+}$}
\newcommand{\kam}{K$^{-}$}

\newcommand{\Dzero}{\ensuremath{\mathrm{D^{0}}}}
\newcommand{\Dzerobar}{\ensuremath{\mathrm{\overline{D^{0}}}}}

\newcommand{\Dstar}{\ensuremath{\mathrm{D^{*\pm}}}}

\newcommand{\gr}[1]{{\foreignlanguage{greek}{#1}}} 
\newcommand{\particle}[1]{\textnormal{#1}} 
\newcommand{\partgr}[1]{{\particle{\gr{#1}}}} 
\newcommand{\pion}{\partgr{p}}

\newcommand{\pip}{\ensuremath{\pion^{+}}}
\newcommand{\pim}{\ensuremath{\pion^{-}}}

\newcommand{\pp}{pp}
\newcommand{\pPb}{p--Pb}
\newcommand{\PbPb}{Pb--Pb}

\newcommand{\GeVc}{\ensuremath{{\rm GeV/}c}}
\newcommand{\gevc}{\ensuremath{{\rm GeV/}c}}
\newcommand{\GeVcsq}{\ensuremath{{\rm GeV/}c^2}}
\newcommand{\tev}{\ensuremath{{\rm TeV}}}
\newcommand{\gev}{\ensuremath{{\rm GeV}}}
\newcommand{\micron}{\partgr{m}m}

\newcommand{\s}{\ensuremath{\sqrt{s}}}

\newcommand{\pt}{\ensuremath{p_{\rm T}}}
\newcommand{\dedx}{d$E$/d$x$}

\newcommand{\zpar}{\ensuremath{z_{||}^{\mathrm{ch}}}}
\newcommand{\zpargen}{\ensuremath{z_{\mathrm{||,gen}}^{\mathrm{ch}}}}
\newcommand{\zpardet}{\ensuremath{z_{\mathrm{||,det}}^{\mathrm{ch}}}}
\newcommand{\ptchjet}{\ensuremath{p_{\mathrm{T,jet}}^{\mathrm{ch}}}}
\newcommand{\ptjet}{\ensuremath{p_{\mathrm{T,jet}}}}
\newcommand{\ptchjetgen}{\ensuremath{p_{\mathrm{T,gen\,jet}}^{\mathrm{ch}}}}

\newcommand{\ptchjetdet}{\ensuremath{p_{\mathrm{T,det\,jet}}^{\mathrm{ch}}}}
\newcommand{\ptd}{\ensuremath{p_{\mathrm{T,D}}}}

\newcommand{\antikt}{anti-\ensuremath{k_{\mathrm{T}}}}

%% file: content.tex
\input{introduction}

\input{apparatus}

\input{analysis}

\input{corrections}

\input{sysUncertainties}

\input{results}

\input{conclusions}

%% file: introduction.tex
\section{Introduction}
\label{sec:introduction}
The study of heavy-flavour production in high-energy interactions provides important tests for Quantum-Chromodynamics (QCD) 
calculations~\cite{Collins:1989gx, Catani:1990eg, Andronic:2015wma}. 
The transverse-momentum ($\pt$)-differential production
cross section of D mesons from charm-quark fragmentation (referred to as ``prompt'' D mesons) was measured
in proton--proton (\pp) and p$\overline{\rm p}$ collisions at several center-of-mass energies, from $\sqrt{s}=0.2~\tev$ at RHIC up to the energies of Tevatron ($\sqrt{s}=1.96~\tev$) and the LHC ($\sqrt{s}=13~\tev$)~\cite{CDF:2003a,STAR:2012a,ATLAS:2015a,ALICE:2012d,ALICE:2016a,ALICE:2017c,LHCb:2013a,LHCb:2016a,LHCb:2016b}. 
The data are described reasonably well by calculations based on perturbative QCD (pQCD) that rely either on the collinear-factorisation approach, like
FONLL~\cite{Cacciari:1998,Cacciari:2001,Cacciari:2012b} and GM-VFNS~\cite{Kniehl:2012a}, or on the $k_{\rm T}$-factorisation 
approach~\cite{Maciula:2013a}. In comparison to single-particle measurements, the reconstruction of jets containing charm hadrons allows for more 
differential studies to characterise the heavy quark production and fragmentation.
A relevant observable 
is the fraction ($z_{||}$) of the jet momentum ($\vec{p}_{\rm jet}$) carried by the D meson along the jet axis direction:
\begin{equation}   
\label{eq:zpar}
z_{||}=\frac{\boldsymbol{\vec{p}}_{\rm jet} \cdot \boldsymbol{\vec{p}}_{\rm D}}{\boldsymbol{\vec{p}}_{\rm jet} \cdot \boldsymbol{\vec{p}}_{\rm jet}},
\end{equation}
where $\vec{p}_{\rm D}$ is the D-meson momentum.

Pioneering measurements of charm jets were performed at the CERN SPS~\cite{UA1:1990a} and at the Tevatron~\cite{CDF:1990,CDF:2004}.
The STAR experiment at RHIC measured the $\Dstar$-meson production in jets in \pp\ collisions at $\s=200~\gev$~\cite{STAR:2009a}.
The jets were measured in the interval $8<\ptjet<20~\gevc$. The yield at low $z_{||}$ values is higher than 
that obtained with a Monte Carlo simulation performed with \mbox{PYTHIA~6}~\cite{Sjostrand:2006a} using only 
the direct charm flavour creation processes, ${\rm gg} \rightarrow {\rm c{\overline{c}}}$ and 
${\rm q{\overline{q}}} \rightarrow {\rm c{\overline{c}}}$. This suggests that higher order processes
(gluon splitting, flavour excitation) are not negligible in the charm production at RHIC energies. 
In a more recent analysis, the PHENIX collaboration measured azimuthal correlations of charm and bottom hadrons in their semi-leptonic decays
using unlike- and like-sign muon pairs~\cite{PHENIX:2018a}. Overall 
they found good agreement with a PYTHIA~6~\cite{Sjostrand:2006a} simulation. Through a Bayesian 
analysis based on PYTHIA~6 templates, the PHENIX collaboration found that while leading order pair creation is dominant
for bottom production, higher order processes dominate for charm one. 

At the LHC, the analysis of the angular correlations of b-hadron 
decay vertices, measured by CMS~\cite{CMS:2011d}, indicated that the collinear region, where the contributions 
of gluon splitting processes are expected to be large, is not adequately described by PYTHIA~6 nor by
predictions based on Next-To-Leading (NLO) order QCD calculations. The ATLAS experiment
measured the $\Dstar$-meson production in jets in pp collisions at $\sqrt{s}=7~\tev$~\cite{ATLAS:2012d}, 
finding that the $z_{||}$ distribution differs from expectations of 
PYTHIA~6, HERWIG~6~\cite{Corcella:2000a,Corcella:2002a} and 
POWHEG~\cite{Nason:2004a,Frixione:2007a,Alioli:2010a,Alioli:2010b} event generators, both 
in overall normalization and shape, with data displaying a higher probability for
low $z_{||}$ values and a steeper decrease towards $z_{||}=1$. The discrepancy between data and generator expectations is
maximum in the lowest jet $\pt$ interval, $25<\pt<30~\gevc$. 
The ATLAS data are well described in a recent global QCD analysis of fragmentation functions based on the ZM-VFNS~\cite{Thorne:2001a} scheme,
in which the in-jet fragmentation data were combined with previous D-meson measurements in a global fit~\cite{Anderle:2017a}.
This global QCD analysis evidences the importance of in-jet fragmentation data in order to pin down the otherwise largely unconstrained momentum fraction dependence of the gluon fragmentation function.

In this paper, we report the first ALICE measurements of the $\Dzero$-meson tagged track-based jet $\pt$-differential cross section in \pp\ collisions at $\sqrt{s}=7~\tev$  and of the \Dzero-meson \zpar\ distribution. The \zpar\ is defined as in Eq.~\ref{eq:zpar} but using the momenta of the track-based jet $\boldsymbol{\vec{p}}_{\rm jet}^{\rm ch}$.
With track-based jets we indicate jets reconstructed with only their 
charged-particle constituents~\cite{Thaler:2013a}. As described in Section~\ref{sec:apparatus}, the excellent low- and intermediate-momentum
tracking capabilities of the ALICE apparatus allow the measurement of jets at very low $\pt$, particularly in the charged jet transverse momentum range $5<\ptchjet<30~\gevc$ considered in 
this paper. This kinematic region is still largely unconstrained by previous measurements.

The measurements reported in this paper are also important to define a \pp\ reference baseline 
for future measurements in \PbPb\ and \pPb\ collisions at the LHC. Charm
quarks, interacting with the constituents of the Quark-Gluon Plasma formed in these collisions, lose energy via both radiative and collisional processes, as 
evidenced by the strong suppression of high-$\pt$ D-meson production measured by ALICE~\cite{ALICE:2012f, ALICE:2015d, ALICE:2016b} 
and CMS~\cite{CMS:2017a}. Contrary to single particles, jets allow one to capture more details of the parton shower dynamics in the medium.
In particular, the study of jet substructure, pioneered for QCD studies and beyond standard model searches \cite{Asquith:2018igt}, can be important to investigate the microscopic properties of hadronic matter at high densities and temperatures~\cite{Sirunyan:2017bsd,Andrews:2018,Acharya:2018uvf,Sirunyan:2018gct}. 

%
%
%

The paper is structured as follows. Section~\ref{sec:apparatus} describes the components of the ALICE apparatus, the data sample and Monte Carlo simulations used
in the analysis. In Section~\ref{sec:DtaggJetReco}, the analysis procedure to obtain the raw spectrum of \Dzero-meson tagged jets and the \zpar\ distribution 
is outlined. Section~\ref{sec:corrections} describes several corrections that are required to account for the $\Dzero$-meson and jet reconstruction efficiency,
the jet momentum scale and the contribution from \Dzero\ mesons coming from b-hadron decays. The systematic uncertainties affecting
the measurements are reported in Section~\ref{sec:systunc}. The results and physics implications are 
discussed in Section~\ref{sec:results}. Finally, Section~\ref{sec:conclusions} closes the paper with conclusions and future perspectives.

%% file: apparatus.tex
\section{Apparatus and data sample}
\label{sec:apparatus}
The measurements presented in this paper were carried out using data recorded by the ALICE apparatus~\cite{Aamodt:2008zz, ALICE:2014b} in 2010.
ALICE is composed of a central barrel embedded in a $0.5$~T magnetic field parallel to the beam direction 
($z$ axis in the ALICE reference frame) and a set of forward- and backward-rapidity detectors. 
The Inner Tracking System (ITS) and the Time-Projection Chamber (TPC) were used for charged-particle track reconstruction 
and the combined information from the TPC and the Time-Of-Flight (TOF) detectors was used to provide particle identification (PID).
These detectors are located in the central barrel, which has a full azimuthal coverage and a pseudorapidity interval of $|\eta|<0.9$.

The ITS is the closest detector to the interaction point and consists of six cylindrical layers of silicon detectors, using three different technologies:
Silicon Pixel Detectors (SPD), whose radius of the first layer is $3.9$~cm, Silicon Drift Detectors (SDD) and Silicon Strip Detectors (SSD). 
The proximity of the SPD to the interaction point, combined with its high spatial resolution, 
provides a resolution on the track impact parameter with respect to the primary vertex better than $75\:\rm\micron$
for tracks with transverse momentum $\pt >1$~\GeVc.

The TPC consists of a $510$~cm long cylinder with an inner radius of $85$~cm and an outer radius of $250$~cm. 
The detector is divided into two halves at the center by a high voltage electrode that generates a uniform electric field in
the longitudinal direction pointing from the endplates to the center. 
The TPC is filled with a mixture of Ne ($90$\%) and CO$_2$ ($10$\%) gases.
The trajectories of charged particles traversing the TPC volume are reconstructed from the ionisation produced in the gas.
The ALICE apparatus is capable of reconstructing charged-particle tracks down to $\pt=0.15~\gevc$ with a $\pt$-resolution better than $2\%$ up to $\pt=20~\gevc$.

The PID information from the TPC is based on the particle specific ionisation energy loss \dedx\ in the gas.
The TOF provides particle identification based on the time-of-flight of the particle from the interaction point to the hit in the Multi-Gap Resistive Plate Chambers (MRPCs) that compose the detector.
For events with sufficiently large multiplicity, the best estimate of the collision time is obtained from the particle arrival times at the TOF~\cite{ALICE:2016c};
for lower-multiplicity events the collision time is measured by the T0 detector,
which consists of two arrays of Cherenkov counters located at $+350$~cm and $-70$~cm along the beam line.
The combined PID information from both detectors provides up to $3\sigma$ separation power for pions/kaons in the range $0.5 < \pt < 2$~\GeVc~\cite{ALICE:2014d}.

The V0 detector was used for triggering minimum-bias events. The detector consists of two scintillator arrays located 
around the beam pipe on each side of the interaction point
covering the pseudo-rapidity interval $-3.7 < \eta < -1.7$ and $2.8 < \eta < 5.1$, respectively.
The minimum-bias condition is defined by the presence of at least one hit in one of the V0 scintillators or in the SPD.

In the work presented in this paper, \pp\ collisions at $\s = 7$~TeV were analysed.
The sample consists of about $388 \times 10^{6}$ minimum-bias events, corresponding to an integrated luminosity of $\mathcal{L}_{\rm int}= 6.23$~nb$^{-1}$~\cite{ALICE:2013e}. 
Events were selected offline by using the timing information from the V0 and the correlation between
the number of hits and track segments in the SPD detector to remove background due to beam--gas
interactions.
Only events with the primary vertex reconstructed within $|z| < 10$~cm with respect to the center of the detector were used for this analysis. 

Monte Carlo (MC) simulations were employed to calculate corrections as described in Section~\ref{sec:corrections}.
The simulations were performed using PYTHIA 6.4.24~\cite{Sjostrand:2006a} with the Perugia 2011 tune~\cite{Skands:2010a}.
The generated particles were transported through the ALICE apparatus using the GEANT3 transport model \cite{GEANT3-url}. The luminous 
region distribution, the geometry of the apparatus, as well as the conditions of all the ALICE detectors were reproduced in detail
in the simulations.

%% file: analysis.tex
\section{Analysis}
\label{sec:DtaggJetReco}

\subsection{D\texorpdfstring{$\boldsymbol{^0}$}{0}-meson selection}
The \Dzero\ mesons were reconstructed via their hadronic decay $\Dzero \rightarrow {\rm K}^{-}\pip$ (and charge conjugate)
which has a branching ratio of $(3.89 \pm 0.04)$\%~\cite{PDG:2018}. In each event, \Dzero-meson candidates and their decay vertices were constructed from pairs of tracks with opposite charge.
The tracks were required to have $|\eta| < 0.8$, $\pt > 0.3$~\GeVc, at least $70$ 
associated TPC space points (out of a maximum of $159$), $\chi^2/ {\rm ndf} < 4$ in the TPC (where ndf is the number of degrees 
of freedom involved in the tracking procedure), at least one hit in either of the two layers of the SPD and a minimum of $3$ hits in the entire ITS.

The \Dzero-meson selection criteria were established in previously published works by the ALICE Collaboration~\cite{ALICE:2012d, ALICE:2017c}.
\Dzero\ mesons were required to be within the rapidity interval comprised by a fiducial detector acceptance,
$|y| < y_{\rm fid}(\ptd)$, with $y_{\rm fid}(\ptd)$ increasing 
from $0.5$ to $0.8$ in the \Dzero-meson transverse momentum interval
$2 < \ptd < 5$~\GeVc\ and $y_{\rm fid}(\ptd) = 0.8$ for $\ptd > 5$~\GeVc.
Outside of this selection the \Dzero-meson reconstruction efficiency drops rapidly
as a consequence of the detector pseudorapidity acceptance and the kinematic selections applied on the tracks.
 
In order to suppress the combinatorial background, we exploited the specific decay topology of the \Dzero\ mesons.
\Dzero\ mesons have a mean proper decay length $c\tau=123\:\rm\micron$~\cite{PDG:2018}. Their decay vertices are therefore typically 
displaced by a few hundred \micron\ from the primary vertex of the interaction.
The selection requirements were tuned to maximise the
statistical significance of the signal along with good reconstruction efficiency. 
The geometrical selections were based on the
displacement of the tracks from the interaction vertex, the distance between the D-meson decay vertex
and the primary vertex (decay length, $L$) and the pointing of the reconstructed D-meson momentum to
the primary vertex in the laboratory reference frame.

Further reduction of the combinatorial background was achieved by applying PID to the decay track candidates. 
The PID selection is based on the ${\rm d}E/{\rm d}x$ and the time-of-flight signals measured with the TPC and TOF detectors, respectively.
The selection was applied by requiring that the difference between the measured and expected PID signals was below $3\sigma$,
where $\sigma$ is the experimental uncertainty associated with the measured signals.
Based on the PID information, \Dzero-meson candidates were accepted (as \Dzero, $\overline{\rm D^0}$, or both) or rejected, 
according to the compatibility with the ${\rm K}^{\mp}\pion^{\pm}$ final state. 
In the cases where both decay track candidates are found to be compatible with both the kaon and pion hypotheses,
the \Dzero-meson candidate was considered twice in either mass combinations corresponding to one of the two possible final states \kam\pip\ and \kap\pim.
The candidates corresponding to a real \Dzero\ meson but with the wrong daughter particle mass assignment are referred to as \emph{reflections}.
This component of the background was subtracted using Monte Carlo templates as described in Section~\ref{sec:rawYield}.

\subsection{Jet reconstruction and D\texorpdfstring{$\boldsymbol{^0}$}{TEXT}-meson tagging}
\label{sec:jetReco}
For jet reconstruction, looser track selection criteria were employed as compared to those used to identify \Dzero-meson candidates.
The pseudorapidity and momentum acceptance windows were
extended to $|\eta| < 0.9$ and $\pt > 0.15$~\GeVc, respectively. The requirement on the SPD hits was lifted
to increase the track-reconstruction efficiency and improve its uniformity as a function of $\eta$ and azimuthal angle $\varphi$.
Tracks without SPD hits were required to contain at least $3$ hits in the ITS and were
constrained to the primary vertex of the interaction. Tracks without SPD hits 
comprise about $19$\% of the track sample used for jet reconstruction. The track reconstruction 
efficiency obtained with these selection criteria is uniform as a function of $\eta$ and $\varphi$.
As a function of the track transverse momentum, the efficiency is about $70$\% for $\pt=0.2$~\GeVc, 
it approaches its maximum value of $90$\% for $\pt\approx2$~\GeVc\ and then it drops again and reaches a plateau at about $85$\%.
The relative track transverse momentum resolution is better than 3\% in the range $0.15<\pt<40$~\GeVc.

Jet reconstruction was performed with the \antikt\ algorithm~\cite{Cacciari:2008c}, as implemented in the \texttt{FastJet}~\cite{Cacciari:2012a} software package,
with a resolution parameter $R=0.4$ and the \pt\ recombination scheme.
From simple kinematic considerations we evaluated that more than 50\% 
of the \Dzero\ mesons with $\ptd=3$~\GeVc\ have their decay products emitted
at an angle larger than $0.4$~rad with respect to the \Dzero-meson momentum direction.
This fraction approaches zero only for $\ptd>7$~\GeVc.
As a consequence, the decay products of low-momentum \Dzero\ mesons are often found outside of the reconstructed jet cone
that is physically correlated with the \Dzero\ meson.
It follows that the decay products of a single \Dzero-meson candidate may be wrongly associated to two different jet candidates in the jet finding phase.
In order to avoid ambiguities in the charm jet tagging and to improve the jet momentum resolution,
a constraint was applied in the jet finding procedure ensuring that pairs of kaons and pions
identified as the decay products of the same  \Dzero-meson candidate are part of the same jet.
This constraint was implemented by removing the 4-momenta of the decay products of identified \Dzero-meson candidates
from the pool of particle tracks used in the jet finding, and replacing them with the 4-momenta of their respectively associated \Dzero-meson candidates.
Events containing more than one \Dzero-meson
candidate passing all the selection criteria are very rare and amount to approximately $0.9$\% of the events that contain at least one accepted candidate.
In these cases, the jet reconstruction procedure was repeated once for each candidate separately,
i.e.\ when analysing one of the candidates, the decay products of the other candidates were
included in the jet reconstruction as single tracks.
This ensures that the combinatorial background of \kam\pip\ track pairs, which dominates the \Dzero-meson candidates at low \pt, does not influence the reconstruction of signal jets.
Jets containing a \Dzero-meson candidate among their constituents were tagged and retained for the next steps of the analysis. 
Jets with $\ptchjet > 5$~\GeVc\ and $|\eta_{\rm jet}|<0.5$ were accepted. The requirement on the jet pseudorapidity ensures that jets are fully contained in the detector acceptance.
No correction to the reconstructed jet \pt\ was performed to account for the background coming from the underlying event (UE), e.g. via multi-parton interactions (MPI).

\subsection{\Dzero-meson tagged jet yield extraction}
\label{sec:rawYield}
\begin{figure}[tbh]
\centering
\includegraphics[width=.85\textwidth]{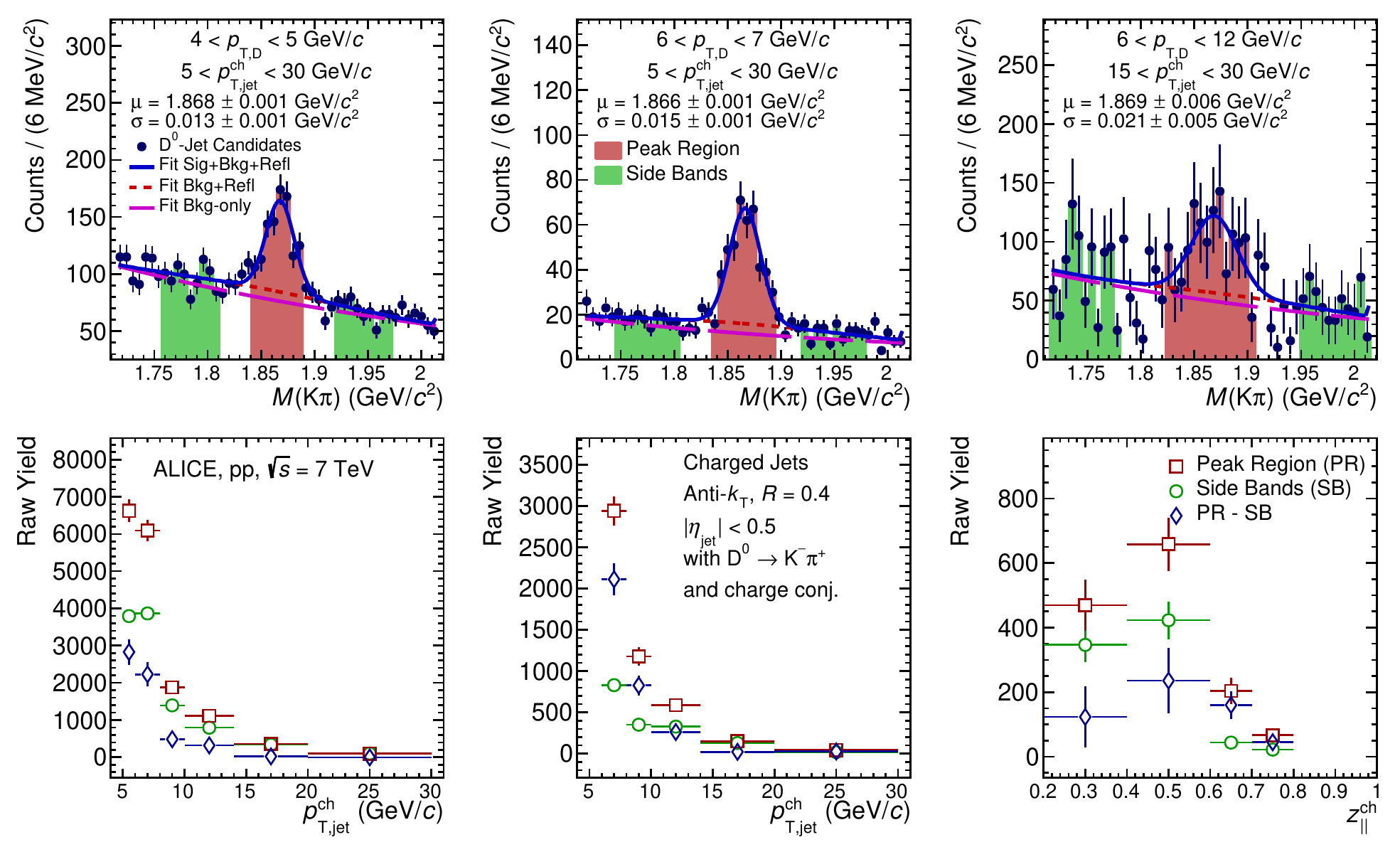}
\caption{(Colour online) Top: Invariant mass distribution of \Dzero-meson tagged jet candidates with $5<\ptchjet<30$~\GeVc\ (left and center) and $15<\ptchjet<30$~\GeVc\ (right).
The \Dzero-meson transverse momenta are required to be in the interval $4<\ptd<5$~\GeVc\ (left), $6<\ptd<7$~\GeVc\ (center) and $6<\ptd<12$~\GeVc\ (right).
The blue solid line represents the total fit function; the background component of the fit function is shown with and without
the reflection component, as a red dotted line and as a magenta dashed line, respectively.
The green and red filled areas correspond to the side-band and peak regions. 
Bottom: Distributions of the \Dzero-meson tagged jet candidates in the peak region (red squares) and the side-band region (green circles) as a function of \ptchjet\ (left and center) 
and \zpar\ (right). The \ptchjet\ and \ptd\ selections are the same as the corresponding top panels.
The blue diamonds show the subtracted distributions corresponding to the raw signals.}
\label{fig:side_band}
\end{figure}

The jet raw yields were extracted with an invariant mass analysis of the \Dzero-meson candidates used to tag the charm jet candidates.
These candidates were first divided in bins of \ptd.
For each interval of \ptd\ the invariant mass distribution was fit with a function composed of a Gaussian function for the signal and an exponential term for the background.
The position $m_{\rm fit}$ and width $\sigma_{\rm fit}$ of the \Dzero-meson invariant mass peak were extracted from the corresponding parameters of the Gaussian component of the fit function.
The top panels in Fig.~\ref{fig:side_band} show the invariant mass distributions of \Dzero-meson candidates in tagged jets in different intervals of \ptd\ and $5<\ptchjet<30$~\GeVc. 
The \Dzero-meson tagged jet candidates were divided in two sub-samples within each \ptd\ interval: (i) the \emph{peak region} corresponding to
candidates with $|m_{\rm inv} - m_{\rm fit}| < 2\,\sigma_{\rm fit}$; and (ii) the \emph{side-band region} corresponding
to candidates with $4\,\sigma_{\rm fit} < |m_{\rm inv} - m_{\rm fit}| < 8\,\sigma_{\rm fit}$.
The filled red and green regions in the plots correspond to the peak and the side-band regions, respectively.

The contribution from residual \Dzero-meson reflections not rejected by PID was accounted for
by including in the fit a template consisting of the sum of two Gaussian functions with centroids and widths fixed to values obtained in the simulation.
The amplitudes were normalized using the signal observed in data,
keeping the ratio of the reflection component over the \Dzero-meson signal fixed to the value obtained in the Monte Carlo simulation.
In the wide invariant mass interval $1.715<m_{\rm inv}<2.015$~\GeVcsq\ used in the fitting procedure, the reflections over signal ratio varies in the range $0.15-0.30$ as a function of \ptd.

The peak region contains a mixture of signal and combinatorial background, while the side-band region is far enough from the \Dzero-meson peak
to be signal-free. The total background $N_{\rm bkg}(\ptd)$ under the peak was extracted from the exponential and reflection components of the invariant-mass fit function by integrating
them in the interval $|m_{\rm inv} - m_{\rm fit}| < 2\,\sigma_{\rm fit}$. 
In order to obtain jet yields as a function of \ptchjet\ or \zpar, distributions as a function of these observables are
constructed for both the peak and side-band regions.
The side-band distribution is scaled such that its 
total integral is equal to $N_{\rm bkg}(\ptd)$ and then subtracted from
the peak-region distribution to obtain the raw yield as a function of \ptchjet:
\begin{equation}
\label{eq:side_band}
N_{\rm raw}(\ptd,~\ptchjet) = N_{\rm PR}(\ptd,~\ptchjet) - \frac{N_{\rm bkg}(\ptd)}{N_{\rm tot, SB}(\ptd)}N_{\rm SB}(\ptd,~\ptchjet),
\end{equation}
where $N_{\rm raw}(\ptd,\ptchjet)$, $N_{\rm PR}(\ptd,\ptchjet)$ and $N_{\rm SB}(\ptd,\ptchjet)$ are the extracted \Dzero-meson-tagged jet raw yield, the peak-region distribution 
and the side-band distribution as a function of \ptchjet\ in each interval of \ptd; $N_{\rm tot, SB}(\ptd)$ is the total integral of the side-bands in each interval of \ptd.
The procedure used to extract the yield as a function of \zpar\ is completely equivalent and is represented by the same Eq.~\ref{eq:side_band} after
replacing \ptchjet\ by \zpar.
The bottom panels of Fig.~\ref{fig:side_band} show the peak-region, side-band (scaled to the total background under the peak)
and subtracted distributions as a function of \ptchjet\ (left and center) and \zpar\ (right).
The distributions are corrected for the reconstruction efficiency and acceptance factor in $|\eta_{\rm jet}| < 0.5$, as described in Section~\ref{sect:efficiency}.

%% file: corrections.tex
\section{Corrections}
\label{sec:corrections}
The \pt-differential cross section of charm jets tagged with \Dzero\ mesons is defined as:
\begin{equation}
\label{eq:cross_section}
\frac{{\rm d}^2\sigma}{{\rm d}\ptchjet{\rm d}\eta_{\rm jet}}(\ptchjet)=\frac{1}{\mathcal{L}_{\rm int}}\frac{1}{\rm BR}\frac{N(\ptchjet)}{\Delta\eta_{\rm jet}\Delta\ptchjet},
\end{equation}
where $N(\ptchjet)$ is the measured yield in each bin of \ptchjet, corrected for the reconstruction efficiency, acceptance and b-hadron feed-down fraction, and unfolded for the detector jet momentum resolution; 
$\Delta\ptchjet$ is the width of the histogram bin; $\Delta\eta_{\rm jet}=1$ is the jet reconstruction acceptance.
Details on the corrections are discussed in the following sections.
The reported yield includes jets containing either \Dzero\ or \Dzerobar\ mesons
with $\ptd>3$~\GeVc.

The distribution of the jet momentum fraction carried by the \Dzero\ meson in the direction of the jet axis (\zpar) is reported as a differential cross section defined as:
\begin{equation}
\label{eq:cross_section_z}
\frac{{\rm d}^3\sigma}{{\rm d}\zpar{\rm d}\ptchjet{\rm d}\eta_{\rm jet}}(\ptchjet,\zpar)=\frac{1}{\mathcal{L}_{\rm int}}\frac{1}{\rm BR}\frac{N(\ptchjet,\zpar)}{\Delta\eta_{\rm jet}\Delta\ptchjet\Delta\zpar}.
\end{equation}
The distribution was measured in the range $0.2<\zpar<1.0$ for
$5<\ptchjet<15$~\GeVc\ and $\ptd>2$~\GeVc\ and in the range $0.4<\zpar<1.0$ for $15<\ptchjet<30$~\GeVc\ and $\ptd>6$~\GeVc.

$N(\ptchjet)$ and $N(\ptchjet,\zpar)$ are normalized such that one count corresponds to a single \Dzero\ meson.
It follows that a jet containing two \Dzero\ mesons will enter this definition twice.
While this choice may seem unnatural for the definition of a jet cross section,
it has the advantage of having a model-independent tagging efficiency.
In fact, if we were to count only once those jets containing two \Dzero\ mesons, then their tagging efficiency would be twice as large.
Then, the overall tagging efficiency would depend on the model-dependent fraction of jets with two \Dzero\ mesons.
\footnote{In real data,
the actual fraction of measured jets with two \Dzero\ mesons is negligible because of the combination
of the low branching ratio of the \Dzero-meson decay channel used in the analysis and the low reconstruction efficiency.}

\subsection{Reconstruction efficiency}
\label{sect:efficiency}
\begin{figure}[tbh]
\centering
\begin{subfigure}[b]{0.45\textwidth}
\includegraphics[width=\textwidth]{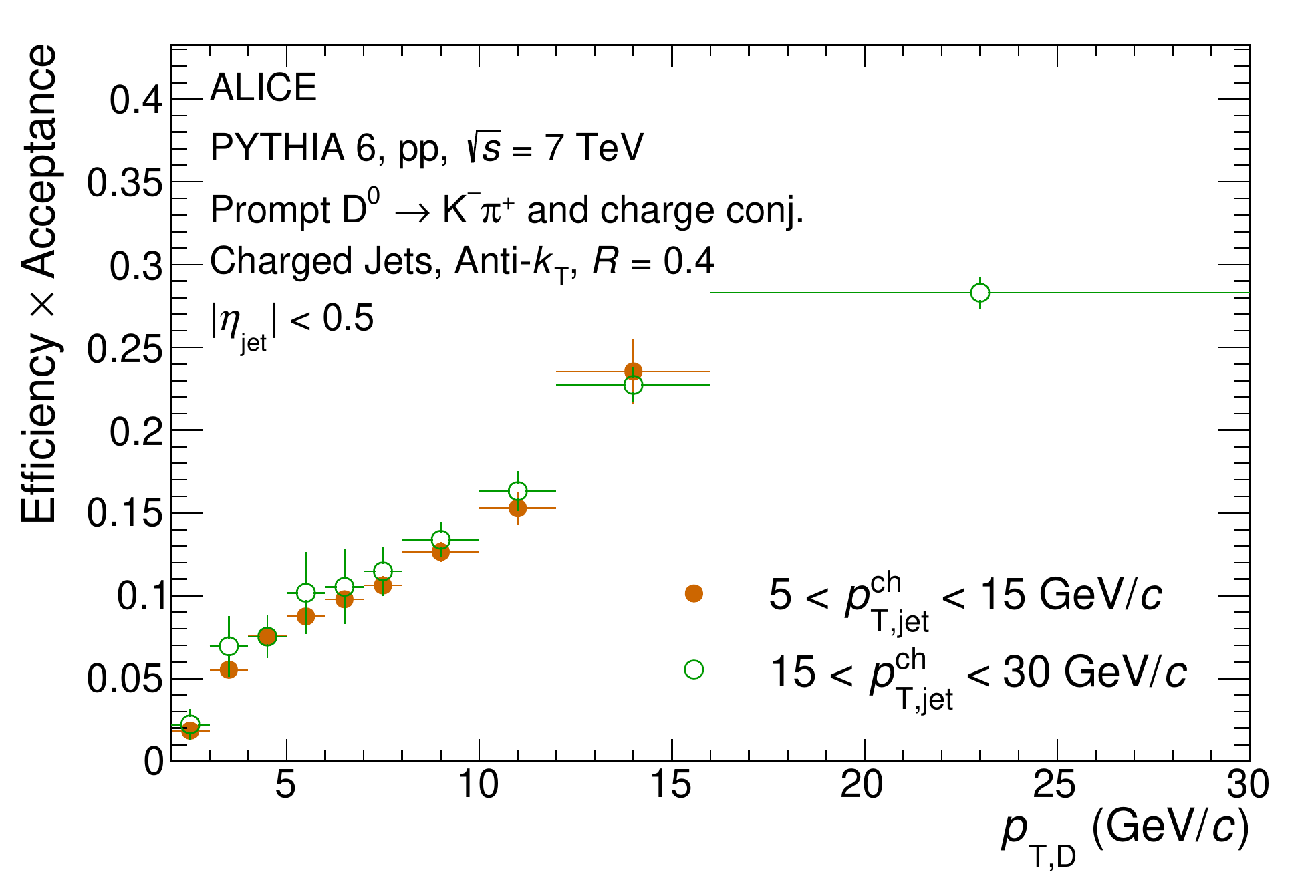}
\end{subfigure}\quad
\begin{subfigure}[b]{0.45\textwidth}
\includegraphics[width=\textwidth]{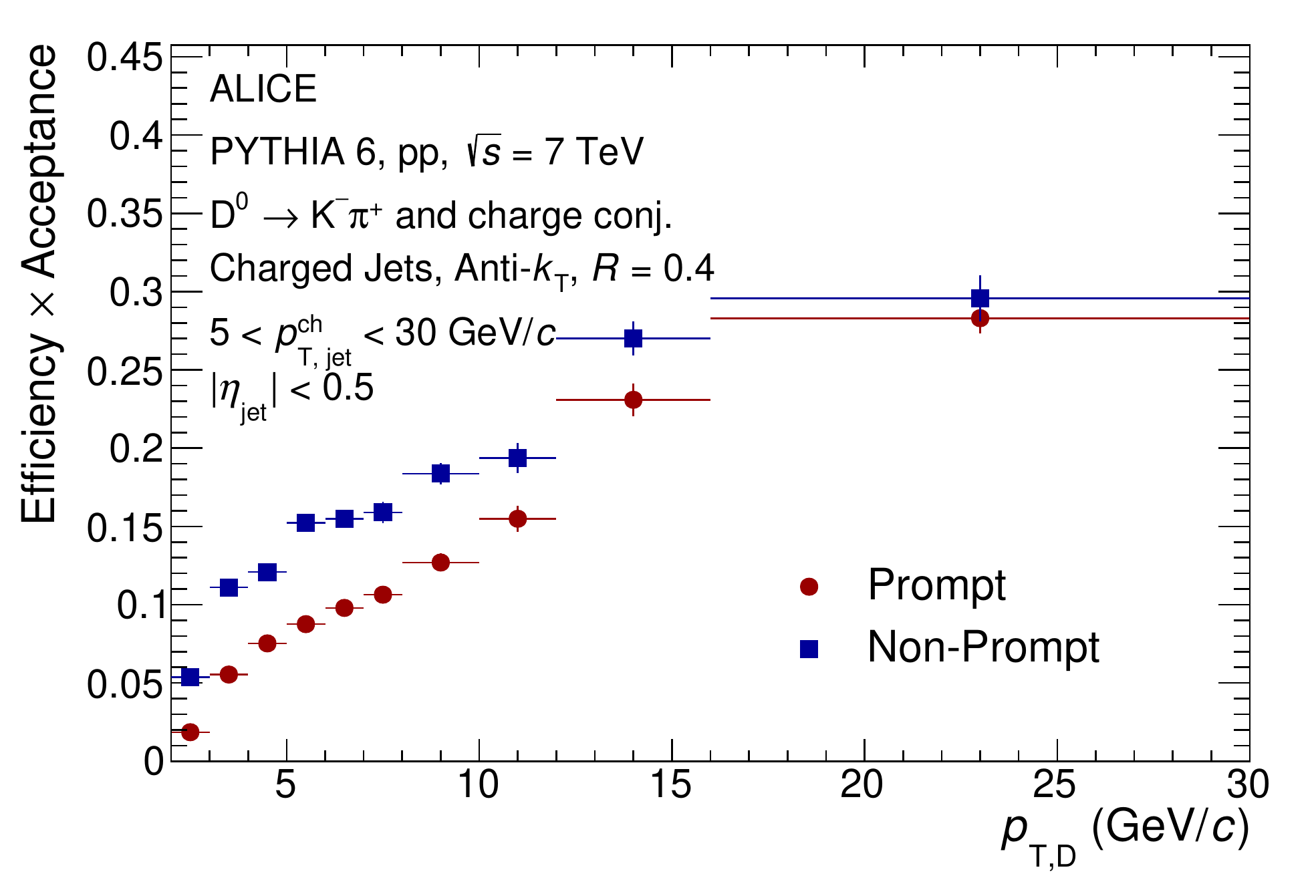}
\end{subfigure}
\caption{Product of acceptance and efficiency of \Dzero-meson jet reconstruction as a function of \ptd. Left: acceptance $\times$ efficiency for prompt \Dzero-meson jets with $5<\ptchjet<15$~\GeVc\ (full circles) and $15<\ptchjet<30$~\GeVc\ (open circles).
Right: prompt (circles) and non-prompt (squares) \Dzero-meson jet acceptance $\times$ efficiency for $5<\ptchjet<30$~\GeVc.}
\label{fig:efficiency}
\end{figure}

The reconstruction efficiency of \Dzero-meson tagged jets depends mainly on the track-reconstruction efficiency
and on the topological selections applied to find the \Dzero-meson candidates.
The efficiency was estimated with a Monte Carlo simulation using the PYTHIA~6 (Perugia 2011)~\cite{Sjostrand:2006a,Skands:2010a} event generator and the GEANT3~\cite{GEANT3-url} transport code.
As shown in Fig.~\ref{fig:efficiency} (left panel), separately for $5<\ptchjet<15$~\GeVc\ and $15<\ptchjet<30$~\GeVc,
the acceptance-times-efficiency is about $6$\% for $\ptd=3$~\GeVc\ and increases rapidly as a function of the \Dzero-meson momentum,
reaching almost $30$\% for $\ptd=30$~\GeVc.

The \ptd\ dependence of the reconstruction efficiency is mainly driven by the topological selections which are much stricter at low \ptd\ in order to suppress the large combinatorial background.
No significant dependence as a function of \ptchjet\ was observed, as the compatibility of the efficiencies for the two \ptchjet\ intervals shows. 

In order to minimize the dependence of the efficiency correction
on the fragmentation model and on the \pt-spectrum shape of the simulated \Dzero-meson tagged jet sample,
the \ptchjet\ distributions were multiplied by the inverse of the efficiency before summing over \ptd:
\begin{equation}
\label{eq:eff_corr}
N_{\rm corr}(\ptchjetdet)=\sum_{\ptd}\frac{N_{\rm raw}(\ptd,\ptchjetdet)}{\epsilon_{\rm P}\left(\ptd\right)},
\end{equation}
where $N_{\rm corr}$ is the efficiency-corrected jet raw yield as a function of reconstructed jet transverse momentum \ptchjetdet,
$N_{\rm raw}(\ptd,\ptchjetdet)$ was defined in Eq.~\ref{eq:side_band},
$\epsilon_{\rm P}\left(\ptd\right)$ is the prompt \Dzero-meson reconstruction efficiency as a function of \ptd.
The sum $\sum_{\ptd}$ is intended over all \ptd\ ranges used in the invariant mass analysis ($3<\ptd<30$~\GeVc).
The same procedure is applied to obtain the yields as a function of \zpar. The corresponding equation is obtained by replacing \ptchjet\ with \zpar\ in Eq.~\ref{eq:eff_corr}.

\subsection{Subtraction of the b-jet contribution}
\label{sect:b_feed_down}

The efficiency of prompt \Dzero-meson tagged jets is lower compared to the efficiency
of those coming from the fragmentation of a beauty quark for which the non-prompt \Dzero-meson is produced by the decay of a beauty hadron. The prompt and non-prompt acceptance and reconstruction efficiency correction factors are compared in Fig.~\ref{fig:efficiency} (right panel).
Due to the longer decay length of beauty hadrons ($c\tau\approx500\:\rm\micron$~\cite{PDG:2018}), some topological selections are more efficient for non-prompt \Dzero\ mesons.
The non-prompt efficiency is higher by about a factor $2$ for $\ptd=3$~\GeVc\ compared to the prompt efficiency. 
The separation between the two efficiencies decreases with \ptd, until they almost converge for $\ptd>15$~\GeVc.

Due to the higher reconstruction efficiency of the non-prompt \Dzero\ mesons, the natural admixture of the prompt and the non-prompt components is biased
towards the non-prompt in a detector- and analysis-specific way. 
In order to simplify comparisons with other experimental results and theoretical calculations,
the fraction of \Dzero-meson tagged jets coming from the fragmentation of b quarks (via the decay of a beauty hadron into a \Dzero) was subtracted as follows.

The non-prompt fraction was estimated with POWHEG interfaced with the PYTHIA~6 (Perugia 2011) Monte Carlo parton shower. The decays of beauty hadrons were turned off in PYTHIA~6, to allow \mbox{EvtGen}~\cite{Lange:2001a} to simulate them. POWHEG was configured with the mass of the b quark $m_{\rm b}=4.75$~\GeVcsq, and the renormalization and factorization scales were kept at the nominal value $\mu_{\rm R}=\mu_{\rm F}=\mu_{0}=\sqrt{\pt^2+m_{\rm b}^2}$. 
The parton distribution function (PDF) was obtained using the LHAPDF~6~\cite{Buckley:2015a} interpolator with
the PDF set CT10nlo~\cite{Lai:2010a}.

\begin{figure}[tbh]
\centering
\begin{subfigure}[b]{0.45\textwidth}
\includegraphics[width=\textwidth]{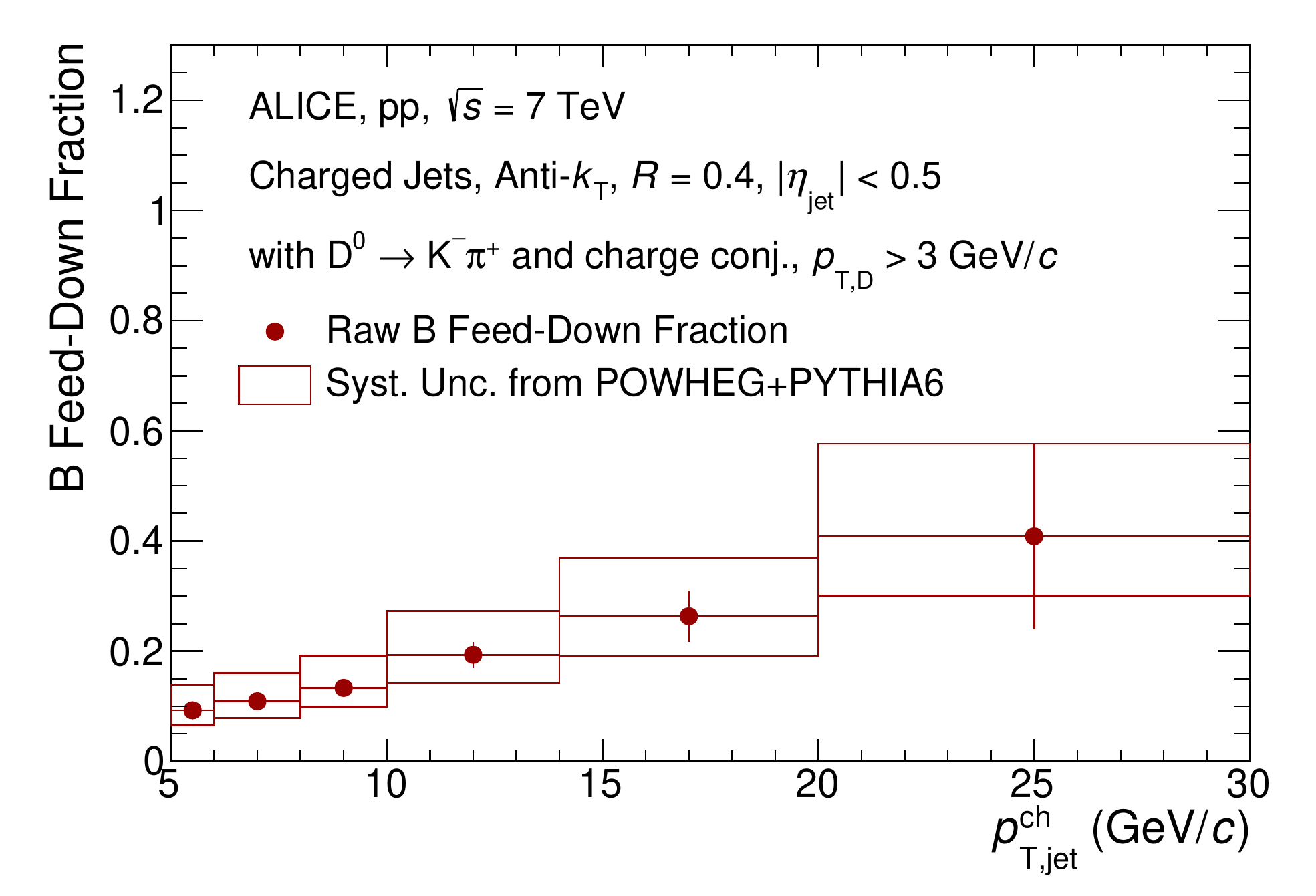}
\end{subfigure}\quad
\begin{subfigure}[b]{0.45\textwidth}
\includegraphics[width=\textwidth]{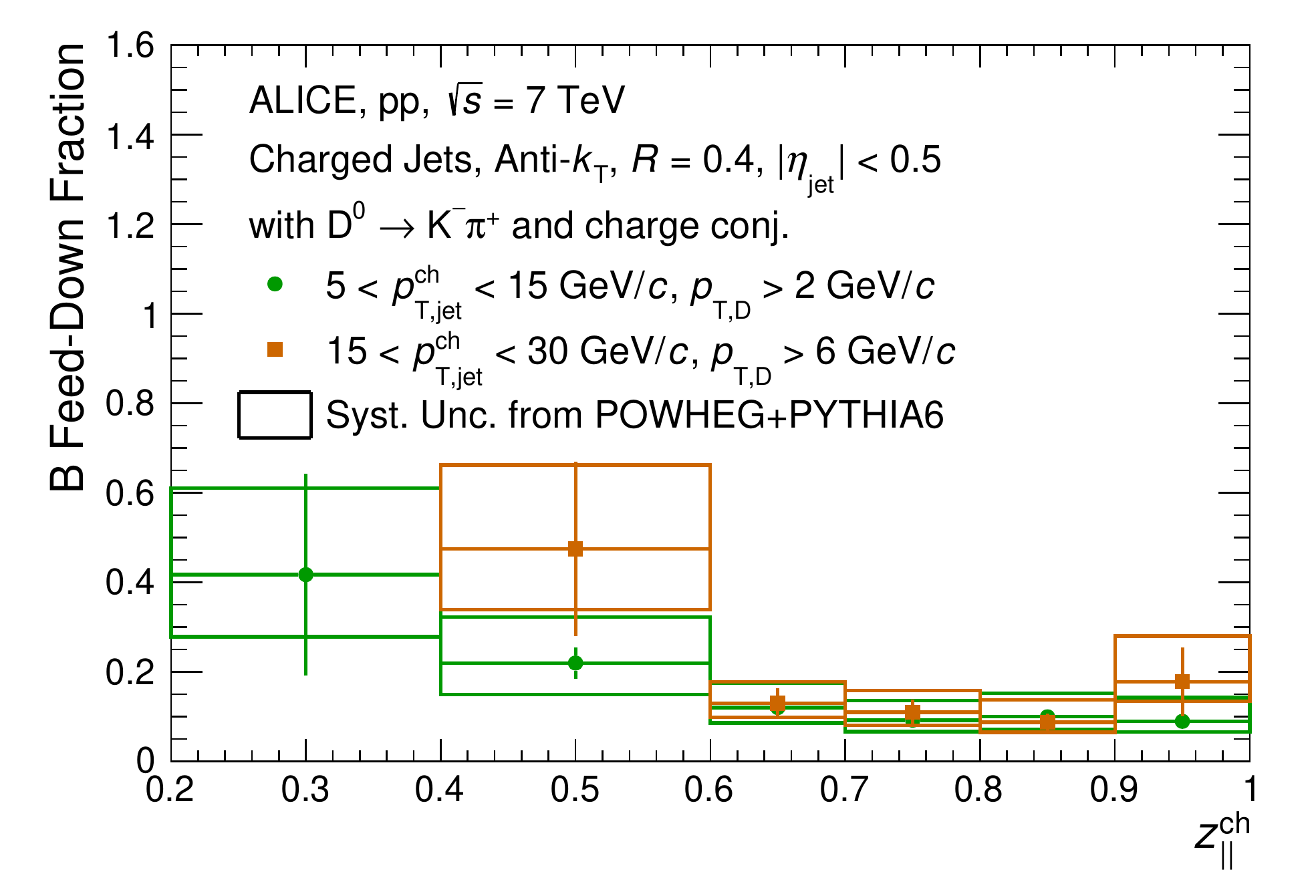}
\end{subfigure}
\caption{b-hadron feed-down fraction of \Dzero-meson tagged jets as a function of \ptchjet\ (left) and \zpar\ (right) in \pp\ collisions at $\s=7$~TeV. On the right, the fraction is shown for $5<\ptchjet<15$~\GeVc\ (circles) and for $15<\ptchjet<30$~\GeVc\ (squares).
The boxes represent the systematic uncertainties, see Section~\ref{sec:systunc} for details.}
\label{fig:b_feed_down}
\end{figure}

The b-hadron feed-down cross sections extracted from the simulation were multiplied by the integrated luminosity of the analyzed data and by the ratio of the non-prompt over the prompt reconstruction efficiencies. A smearing was also applied to account for the detector resolution of the jet momentum.
The b-hadron feed-down fraction was then subtracted from the efficiency-corrected \Dzero-meson tagged jet yield:
\begin{equation}
\label{eq:b_feed_down}
N_{\rm sub}(\ptchjetdet)=N_{\rm corr}(\ptchjetdet)- R_{\rm NP}(\ptchjetdet,\ptchjetgen) \cdot \sum_{\ptd}  \frac{\epsilon_{\rm NP}\left(\ptd\right)}{\epsilon_{\rm P}\left(\ptd\right)} N_{\rm NP}(\ptd,\ptchjetgen),
\end{equation}
where $R_{\rm NP}$ is the matrix representing the \ptchjet\ detector response for non-prompt \Dzero-meson tagged jets (described in more detail in Section~\ref{sec:unfolding});
$\epsilon_{\rm NP}\left(\ptd\right)$ is the reconstruction efficiency of the non-prompt fraction; $N_{\rm NP}(\ptd,\ptchjetgen)$ is the vector corresponding to the b-hadron feed-down yields extracted from the
simulation by multiplying the cross section by the integrated luminosity $\mathcal{L}_{\rm int}$ and discretizing it in bins of \ptd\ and \ptchjetgen.
The sum $\sum_{\ptd}$ is intended over the same \ptd\ ranges used in the signal extraction in data ($3<\ptd<30$~\GeVc\ for the jet \pt-differential cross section).

Figure~\ref{fig:b_feed_down} shows the fraction of non-prompt \Dzero-meson tagged jets as a function of \ptchjet\ (left) and \zpar\ (right).
The estimated fraction shows a steady linear increase as a function of \ptchjet. The dependence on \zpar\ is weak and it appears to decrease only slightly for $\zpar>0.6$.

\subsection{Unfolding}
\label{sec:unfolding}
\begin{figure}[tbh]
\centering
\begin{subfigure}[b]{0.45\textwidth}
\includegraphics[width=\textwidth]{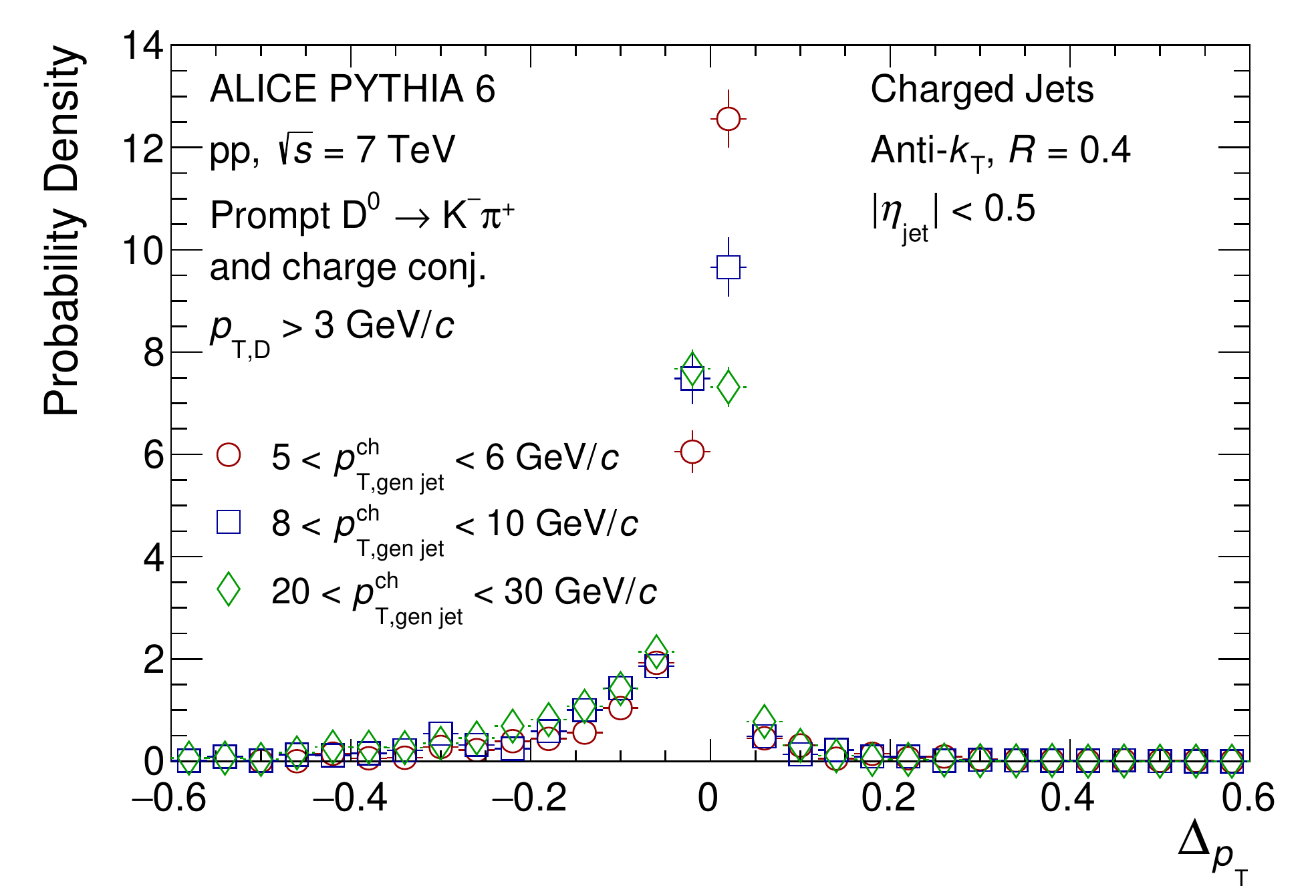}
\end{subfigure}\quad
\begin{subfigure}[b]{0.45\textwidth}
\includegraphics[width=\textwidth]{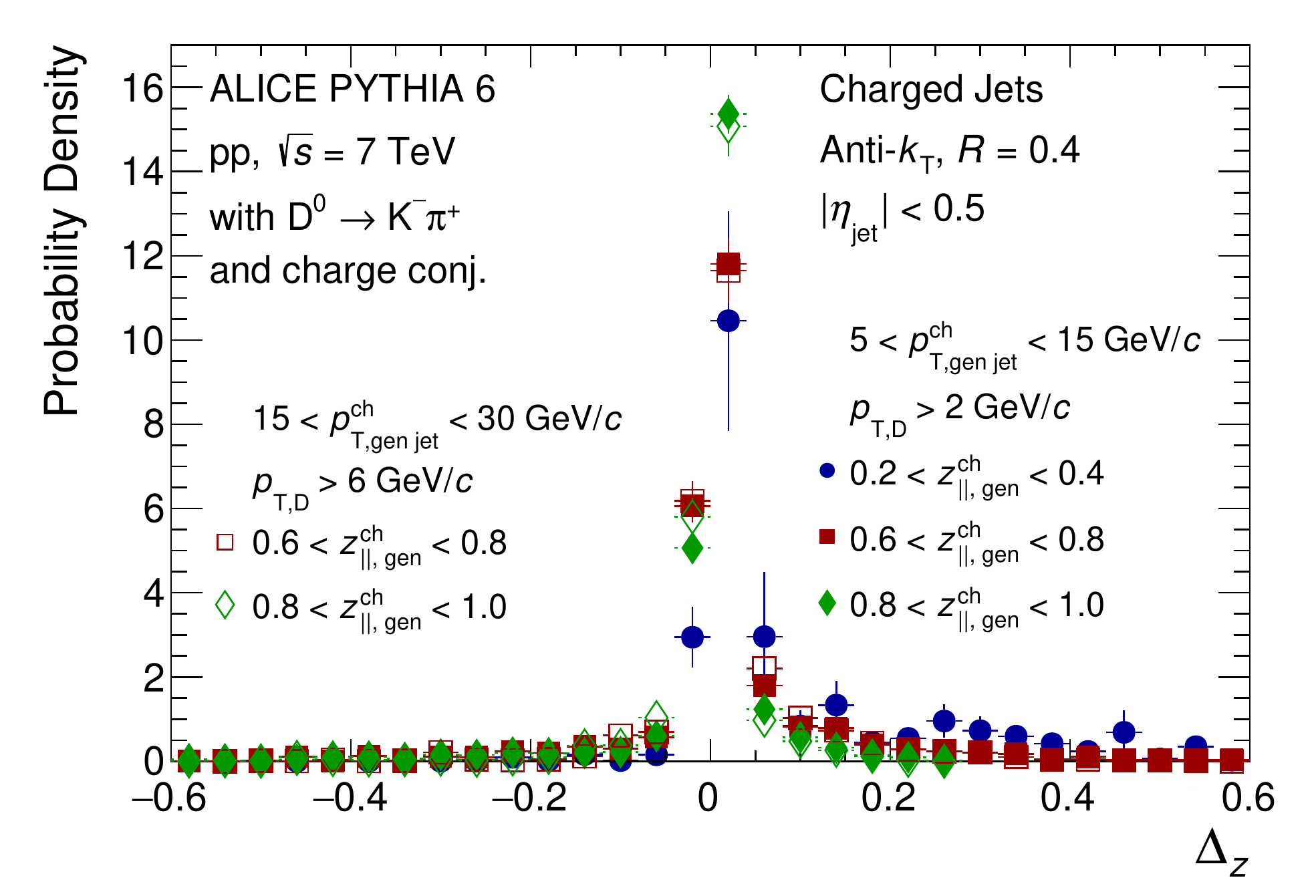}
\end{subfigure}
\caption{Probability density distribution of $\Delta_{\pt}$ (left) and $\Delta_{z}$ (right) for \Dzero-meson tagged jets in \pp\ collisions at $\s=7$~TeV. 
Left: $\Delta_{\pt}$ is shown for $5<\ptchjetgen<6$~\GeVc\ (circles), $8<\ptchjetgen<10$~\GeVc\ (squares) and $20<\ptchjetgen<30$~\GeVc\ (diamonds).
Right: $\Delta_{z}$ is shown for $0.8<\zpargen<1$ (diamonds), $0.6<\zpargen<0.8$ (squares), $0.2<\zpargen<0.4$ (circles); solid markers are used for $5<\ptchjetgen<15$~\GeVc, while open markers represent $15<\ptchjetgen<30$~\GeVc (the data set for $0.2<\zpargen<0.4$ is omitted for the latter).}
\label{fig:det_res}
\end{figure}

The reconstructed jet momentum is affected by the finite detector resolution.
The main factor impacting the jet momentum resolution is the track-reconstruction efficiency,
which causes an average negative shift and a smearing of the reconstructed jet momentum compared to the true jet momentum.
The detector resolution was quantified with the same Monte Carlo simulation used to estimate the efficiency.
It was verified that the simulation is able to reproduce at the detector level the main features of the data, such as jet and \Dzero-meson \pt\ distributions,
and the average number of jet constituents.
\Dzero-meson tagged jets at the detector level were uniquely matched with the corresponding
jets at the generator level.
The matching criteria are based on the presence of the same \Dzero\ meson, which was followed from the generator level throughout its decay and transport in the detector volume.
The jet transverse momentum resolution can be quantified from the probability density distribution of the relative difference ($\Delta_{\pt}$) between the reconstructed jet transverse momentum $\ptchjetdet$
and the generated jet transverse momentum $\ptchjetgen$:
\begin{equation}
\label{eq:jet_pt_res}
\Delta_{\pt}=(\ptchjetdet-\ptchjetgen)/\ptchjetgen.
\end{equation}
A similar quantity is defined for the jet momentum fraction carried by the \Dzero:
\begin{equation}
\label{eq:jet_z_res}
\Delta_{z}=(\zpardet-\zpargen)/\zpargen.
\end{equation}
Figure~\ref{fig:det_res} shows the probability density distributions of $\Delta_{\pt}$ (left) and $\Delta_{z}$ (right) for a selection of \ptchjet\ and \zpar\ ranges.

The mean relative shift of the reconstructed jet momentum varies monotonically from $-2$\% for $\ptchjetgen=5$~\GeVc\ to $-7$\% for $\ptchjetgen=30$~\GeVc.
The resolution, defined as the standard deviation from the mean of the probability density distribution, also varies monotonically as a function of \ptchjetgen\ from $10$\% to $15$\%.
The resolution is slightly better compared to the inclusive jet measurement performed on the same dataset with similar techniques~\cite{ALICE:2015e}.
This difference can be ascribed to the requirement of the presence of a \Dzero\ meson with $\ptd>3$~\GeVc\ in the jet.

Similarly, the mean shift of \zpar\ was found to reach its maximum of $14$\% ($11$\%) for $\zpargen=0.2$ and $5<\ptchjetgen<15$~\GeVc\ ($15<\ptchjetgen<30$~\GeVc), decreasing monotonically with increasing \zpargen\
and approaching zero in the limit $\zpargen=1$ for both ranges of \ptchjetgen. The resolution varies in the range 8--25\% (7--20\%) for  $5<\ptchjetgen<15$~\GeVc\ ($15<\ptchjetgen<30$~\GeVc), 
where the best resolution is obtained for larger values of \zpargen.

The finite detector resolution modifies the dependence of the measured yields as a function of \ptchjet\ and \zpar.
The relationship between the raw and the generated yields can be written as:
\begin{equation}
\label{eq:unfolding_jetpt}
N_{\rm det}(\ptchjetdet)=R_{\rm P}(\ptchjetdet,\ptchjetgen) \cdot N_{\rm gen}(\ptchjetgen),
\end{equation}
\begin{equation}
\label{eq:unfolding_zpar}
N_{\rm det}(\zpardet)=R_{\rm P}(\zpardet,\zpargen) \cdot N_{\rm gen}(\zpargen),
\end{equation}
where $R_{\rm P}$ is the matrix representing the \ptchjet\ detector response for prompt \Dzero-meson tagged jets;
$N_{\rm det}$ and $N_{\rm gen}$ are the vectors corresponding to the measured and generated yields in bins of either \ptchjet\ or \zpar.

The effects of the limited detector resolution discussed above were corrected through an unfolding procedure.
The measured distributions $N_{\rm sub}$ were unfolded using an iterative approach based on Bayes' theorem~\cite{Dagostini:1995a}.
The iterative unfolding algorithm successfully converged after three iterations.
The $N_{\rm sub}(\ptchjet)$ distribution and the two $N_{\rm sub}(\zpar)$ distributions for $5<\ptchjet<15$~\GeVc\ and for $15<\ptchjet<30$~\GeVc\ were
each unfolded separately with their corresponding detector response matrices.
For the \zpar\ distributions an additional correction, based on the same PYTHIA~6 + GEANT3 simulation,
was applied to account for the effect of the detector resolution on \ptchjet,
which causes jet candidates to fall in or out of the considered \ptchjet\ intervals.
This correction is about $+15$\% for jets tagged with \Dzero\ mesons with $\ptd<5$~\GeVc\ and negligible for $\ptd>5$~\GeVc.
The same histogram binning was used for the measured and the unfolded distributions.
Underflow ($\ptchjet<5$~\GeVc) and overflow ($\ptchjet>30$~\GeVc) bins were excluded from the \ptchjetdet\ axis of the response matrix,
but were kept as degrees of freedom in the \ptchjetgen\ axis that could be populated according to the probabilities mapped by the response matrix.
The same applies, only for the underflow bin, for the \zpar\ distributions ($\zpar\leq1$ by construction).
The overall unfolding corrections on the yields are:
between $+2$\% and $+14$\% for the \ptchjet\ distribution;
between $-6$\% and $+5$\% for the \zpar\ distribution with $5<\ptchjet<15$~\GeVc;
between $-30$\% and $+10$\% for the \zpar\ distribution with $15<\ptchjet<30$~\GeVc.
In all cases, the unfolding correction is smaller than the statistical uncertainties.

%% file: sysUncertainties.tex
\section{Systematic uncertainties}
\label{sec:systunc}
The relative systematic uncertainties on the \pt-differential \Dzero-meson tagged jet cross section 
and on the \zpar\ distributions for $5<\ptchjet<15$~\GeVc\ and $15<\ptchjet<30$~\GeVc\ are summarized 
in Tables~\ref{tab:UncSum},~\ref{tab:UncSum_z_low} and~\ref{tab:UncSum_z_high}, respectively. 
In the following, each source of systematic uncertainty is discussed.

The uncertainty on the track-reconstruction efficiency affects our measurement via an uncertainty on the reconstruction efficiency of the \Dzero\ meson and an uncertainty on the jet momentum resolution.
For the \Dzero-meson reconstruction efficiency, a \pt-independent systematic uncertainty of $4$\% was assigned 
based on the \Dzero-meson studies in~\cite{ALICE:2017c}.
The relative systematic uncertainty on the track-reconstruction efficiency for the set of tracks used for jet reconstruction was estimated to be 5\% in~\cite{ALICE:2015e}.
Therefore, the detector response matrix was modified by randomly rejecting $5$\% of the tracks reconstructed in the detector simulation.
The jet \pt\ distribution and \zpar\ distributions were unfolded using this modified matrix and compared with the distributions unfolded with the nominal matrix.
The relative differences were found to be less than 7\% in most cases.
The uncertainty on the track momentum resolution was determined to have a negligible effect on the jet momentum resolution.
Tracks of charged particles produced in the decays of neutral strange hadrons or in secondary interactions with the detector material (including photon conversions)
are largely suppressed by the track selection criteria used in the jet finding. The residual contamination is reproduced fairly well by the Monte Carlo simulation used to estimate the detector response:
this residual contamination is corrected for in the unfolding procedure.
However, PYTHIA~6 does not adequately reproduce the strange particle production~\cite{ALICE:2015h}. 
An uncertainty of about $0.5$\% on the jet momentum arises from this~\cite{ALICE:2018b}, which causes an uncertainty of about $2$\% on the \pt- and \zpar-differential yields.
A possible influence of the simulated \pt-spectrum shape of charm jets on the \Dzero-meson reconstruction efficiency was investigated by re-calculating 
the corrections using an alternative \pt-spectrum shape obtained from an independent simulation in which POWHEG replaced PYTHIA~6 for the generation of the hard scattering.
The effect on the final results was found to be negligible.

Discrepancies between simulation and data that affect the \Dzero-meson reconstruction and selection efficiency introduce a systematic uncertainty.
For example, the selections based on the displacement of the decay vertex from the collision point are sensitive to the resolution
on the track impact parameter with respect to the primary vertex;
residual misalignment of the silicon pixel detector can also introduce irreducible differences between data and simulation.
The systematic uncertainty arising from these discrepancies was determined by repeating the analysis with different sets of selection criteria.
In the D-meson cross-section analysis~\cite{ALICE:2017c}, this uncertainty was estimated to be $5$\%.
Since the uncertainty was also found to depend weakly on the \ptd, for the range considered in this analysis,
and the \Dzero-meson tagged jet reconstruction efficiency does not depend significantly on the \ptchjet\
(see Fig.~\ref{fig:efficiency}), the same value of $5$\% was assigned as uncertainty on the yield for this measurement.

The systematic uncertainties on the raw yield extraction were estimated 
by repeating the fitting procedure of the invariant mass distributions several times with different fit conditions.
The tests included the following:
(i) variations of the upper and lower limits of the fit range;
(ii) variations of the invariant-mass distribution bin width;
(iii) background fit function (default exponential replaced by first and second order polynomial functions);
(iv) mean and/or $\sigma$ parameters of the Gaussian function fixed to the values expected from Monte Carlo simulations.
The root-mean-square of the differences between signal-yield distributions obtained from the various trials was taken as the systematic uncertainty. 
An additional systematic uncertainty was assigned by varying the assumed relative contribution 
of the \Dzero-meson reflections to the signal by $\pm$~50\%. 
The total uncertainty on the \pt-differential jet cross section varies between 4--15\% and rises with \ptchjet.
The uncertainty on the \zpar\ distributions is 2--23\% with higher values for the two lowest \zpar\ intervals.

In order to estimate the systematic uncertainty on the simulation used to subtract the b-hadron feed-down, the following parameters of the POWHEG simulation were varied:
the b-quark mass, the perturbative scales ($\mu_{\rm R, \rm F}$) and the PDF.
The systematic uncertainties were obtained by taking the largest upward and downward variations in the final yields.
In addition, another source of uncertainty was taken into consideration by using the PYTHIA~6 decayer instead of EvtGen to decay the beauty hadrons.
The b-hadron feed-down fractions with their systematic uncertainties are shown in Fig.~\ref{fig:b_feed_down} as a function of \ptchjet\ and \zpar.
This yields systematic uncertainties between 5--23\% increasing (decreasing) with \ptchjet\ (\zpar).

It was verified that after the $3^{\mathrm{rd}}$ Bayesian iteration the unfolding procedure converges and subsequent iterations do not differ significantly from the previous one.
In addition, the prior spectrum used as initial guess was varied in a wide range, by using power-law functions with exponents differing by up to $4$ units from each other.
Finally, a different unfolding technique based on the Singular Value Decomposition (SVD) method~\cite{Hocker:1995} was used, 
as well as a simple bin-by-bin correction technique. In all of these tests, the deviations from the nominal result were found to be smaller than the statistical uncertainties of the measurement. 
Therefore, the systematic uncertainties were assigned using a Monte Carlo closure test. In this test, a detector-level simulation of \Dzero-meson tagged jets 
with a statistical precision comparable to our data was unfolded 
using a detector response matrix obtained from a different and larger Monte Carlo sample.
The unfolded result was compared with the generator-level spectrum.
A \pt-independent $5$\% systematic uncertainty was assigned based on the maximum deviations observed between the unfolded results and the truth.

\begin{table}[bth]
\caption{Summary of systematic uncertainties as a function of \ptchjet.}
\label{tab:UncSum}
\begin{center}
\begin{tabular}{lrrrrrr}
Source & \multicolumn{6}{c}{Uncertainty (\%)} \\ \hline
\ptchjet\ (\GeVc) & 5--6 & 6--8 & 8--10 & 10--14 & 14--20 & 20--30\\ \hline
Tracking Eff. (Jet Energy Scale) & 1 & 3 & 4 & 6 & 7 & 8\\
Raw Yield Extraction & 4 & 4 & 4 & 4 & 11 & 15\\
\Dzero\ Reflections & 3 & 2 & 2 & 3 & 5 & 6\\
Feed-down (POWHEG) & 5 & 5 & 7 & 10 & 17 & 21\\
Feed-down (decayer) & 1 & 1 & 1 & 2 & 4 & 6\\
Unfolding & 5 & 5 & 5 & 5 & 5 & 5\\
PID and Topological Selections & 5 & 5 & 5 & 5 & 5 & 5\\
Tracking Eff. (D Meson) & 4 & 4 & 4 & 4 & 4 & 4\\
Secondary Track Contamination & 2 & 2 & 2 & 2 & 2 & 2\\
\hline
Normalization (BR \& lumi) & \multicolumn{6}{c}{3.6} \\
\hline
Total & 12 & 12 & 13 & 16 & 24 & 30\\
\end{tabular}
\end{center}
\end{table}

Finally, the normalization of the \pt-differential cross section is affected 
by uncertainties on the $\Dzero \rightarrow {\rm K} ^{-}\pip$ branching ratio ($1$\%~\cite{PDG:2018})
and on the minimum-bias trigger efficiency ($3.5$\%~\cite{ALICE:2013e}).

The total systematic uncertainties on the cross section were obtained by summing in quadrature the uncertainties estimated for each of the sources outlined above.
They rise slightly with increasing \ptchjet\ and are comparable to the statistical uncertainties,
except for $\ptchjet>20$~\GeVc, where the statistical uncertainty dominates.
Similarly, the \zpar\ distribution for $5<\ptchjet<15$~\GeVc\ is affected by statistical and systematic uncertainties at a comparable level, 
while for $15<\ptchjet<30$~\GeVc\ statistical ones dominate.

\begin{table}[bth]
\caption{Summary of systematic uncertainties as a function of \zpar\ for $5<\ptchjet<15$~\GeVc.}
\label{tab:UncSum_z_low}
\begin{center}
\begin{tabular}{lrrrrrr}
Source & \multicolumn{6}{c}{Uncertainty (\%)} \\ \hline
\zpar\ & 0.2--0.4 & 0.4--0.6 & 0.6--0.7 & 0.7--0.8 & 0.8--0.9 & 0.9--1.0\\ \hline
Tracking Eff. (Jet Energy Scale) & 5 & 4 & 2 & 2 & 2 & 2\\
Raw Yield Extraction & 23 & 17 & 5 & 3 & 2 & 2\\
\Dzero\ Reflections & 9 & 7 & 4 & 3 & 2 & 2\\
Feed-down (POWHEG) & 22 & 17 & 7 & 4 & 4 & 4\\
Feed-down (decayer) & 8 & 5 & 2 & 2 & 3 & 4\\
Unfolding & 5 & 5 & 5 & 5 & 5 & 5\\
PID and Topological Selections & 5 & 5 & 5 & 5 & 5 & 5\\
Tracking Eff. (D Meson) & 4 & 4 & 4 & 4 & 4 & 4\\
Secondary Track Contamination & 2 & 2 & 2 & 2 & 2 & 2\\
\hline
Normalization (BR \& lumi) & \multicolumn{6}{c}{3.6} \\
\hline
Total & 36 & 27 & 13 & 11 & 11 & 11\\
\end{tabular}
\end{center}
\end{table}
    
\begin{table}[bth]
\caption{Summary of systematic uncertainties as a function of \zpar\ for $15<\ptchjet<30$~\GeVc.}
\label{tab:UncSum_z_high}
\begin{center}
\begin{tabular}{lrrrrr}
Source & \multicolumn{5}{c}{Uncertainty (\%)} \\ \hline
\zpar\  & 0.4--0.6 & 0.6--0.7 & 0.7--0.8 & 0.8--0.9 & 0.9--1.0\\ \hline
Tracking Eff. (Jet Energy Scale) & 4 & 3 & 4 & 9 & 13\\
Raw Yield Extraction & 17 & 11 & 5 & 6 & 7\\
\Dzero\ Reflections & 8 & 7 & 2 & 2 & 2\\
Feed-down (POWHEG) & 20 & 14 & 3 & 5 & 6\\
Feed-down (decayer) & 9 & 6 & 4 & 7 & 8\\
Unfolding & 5 & 5 & 5 & 5 & 5\\
PID and Topological Selections & 5 & 5 & 5 & 5 & 5\\
Tracking Eff. (D Meson) & 4 & 4 & 4 & 4 & 4\\
Secondary Track Contamination & 2 & 2 & 2 & 2 & 2\\
\hline
Normalization (BR \& lumi) & \multicolumn{5}{c}{3.6} \\
\hline
Total & 31 & 22 & 12 & 17 & 20\\
\end{tabular}
\end{center}
\end{table}

%% file: results.tex
\section{Results}
\label{sec:results}

\begin{figure}[tbh]
\centering
\begin{subfigure}[b]{0.45\textwidth}
\includegraphics[width=\textwidth]{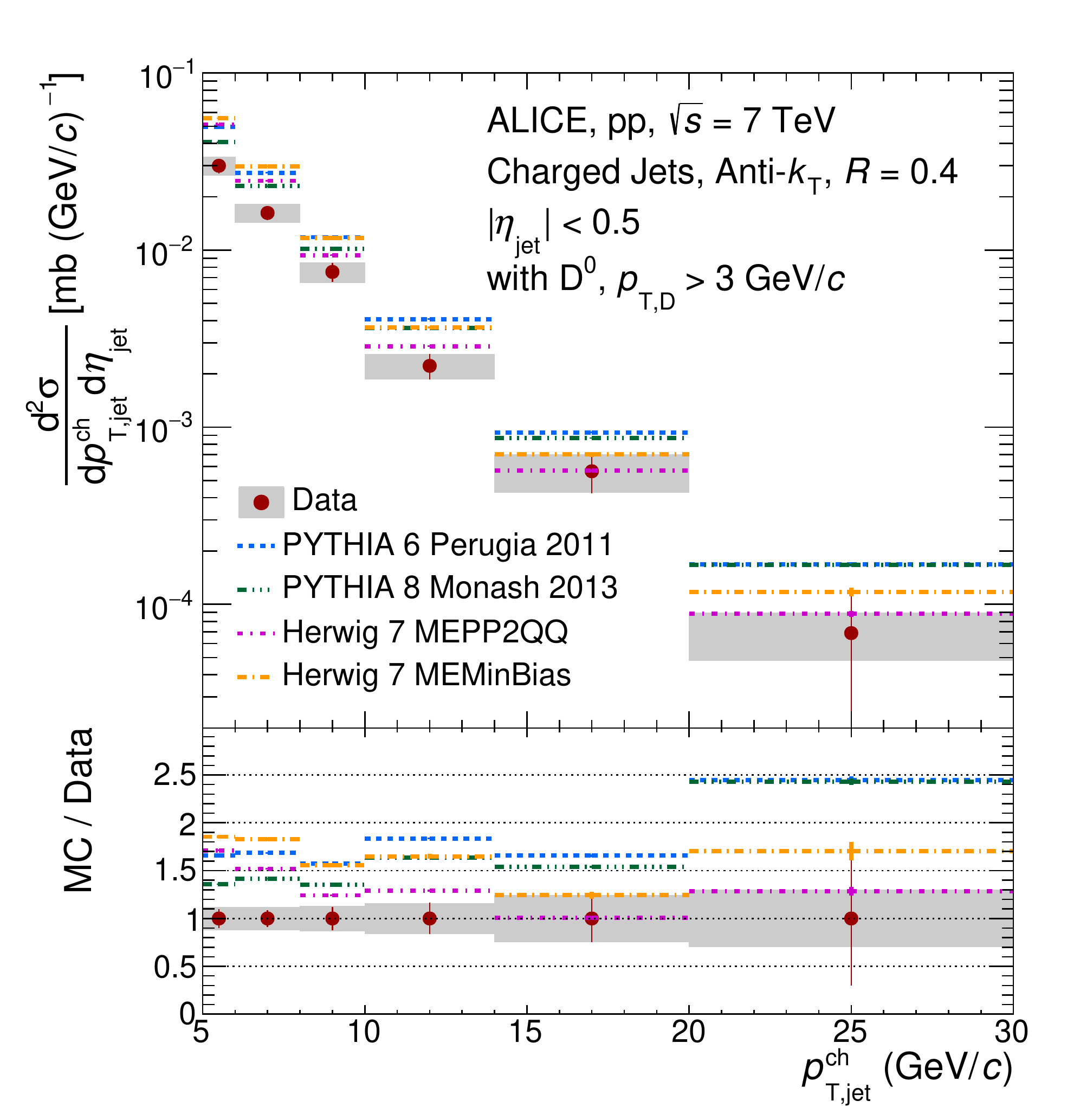}
\end{subfigure}\quad
\begin{subfigure}[b]{0.45\textwidth}
\includegraphics[width=\textwidth]{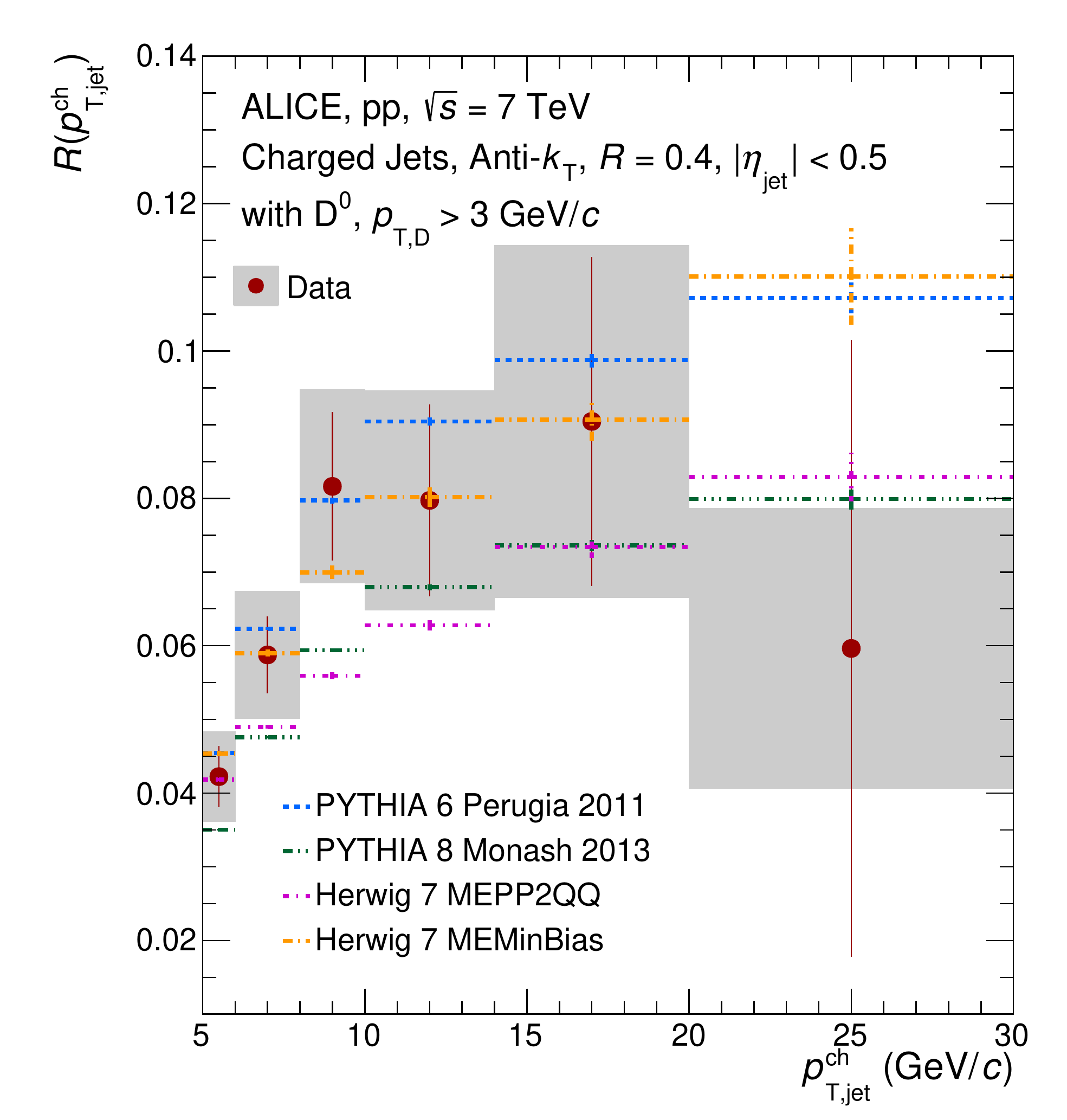}
\end{subfigure}
\caption{(Colour online) \pt-differential cross section of charm jets tagged with \Dzero\ mesons (left) and its ratio to the inclusive jet cross section (right) in \pp\ collisions at $\s=7$~TeV.
The solid red circles show the ALICE data with their systematic uncertainties represented by the grey boxes. 
The measurements are compared with \mbox{PYTHIA~6} Perugia 2011 (blue), \mbox{PYTHIA~8} Monash 2013 (green), Herwig 7 \texttt{MEPP2QQ} (magenta) and \texttt{MEQCDMinBias} (orange).}
\label{fig:cross_section_mc}
\end{figure}

\begin{figure}[tbh]
\centering
\begin{subfigure}[b]{0.45\textwidth}
\includegraphics[width=\textwidth]{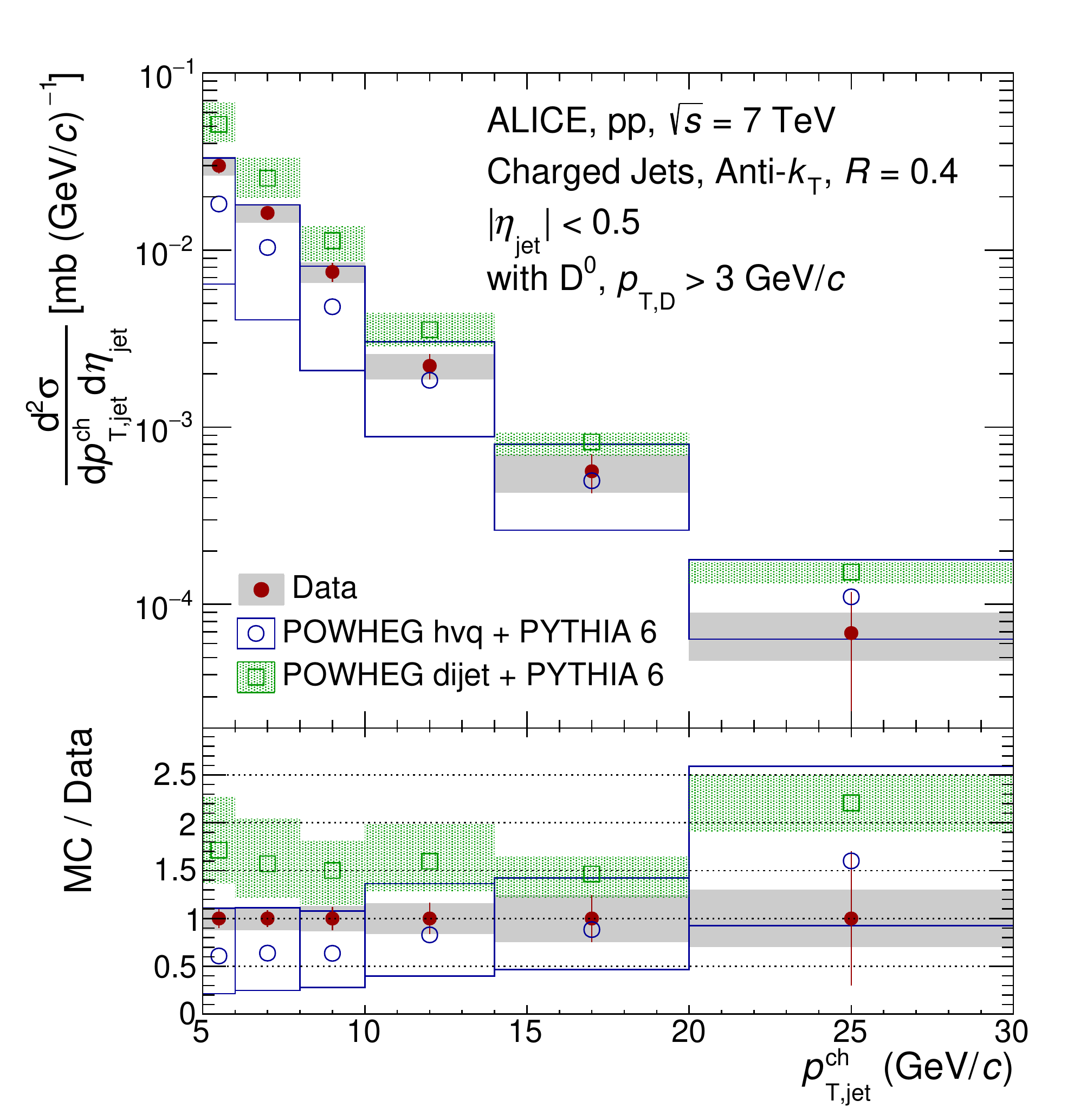}
\end{subfigure}\quad
\begin{subfigure}[b]{0.45\textwidth}
\includegraphics[width=\textwidth]{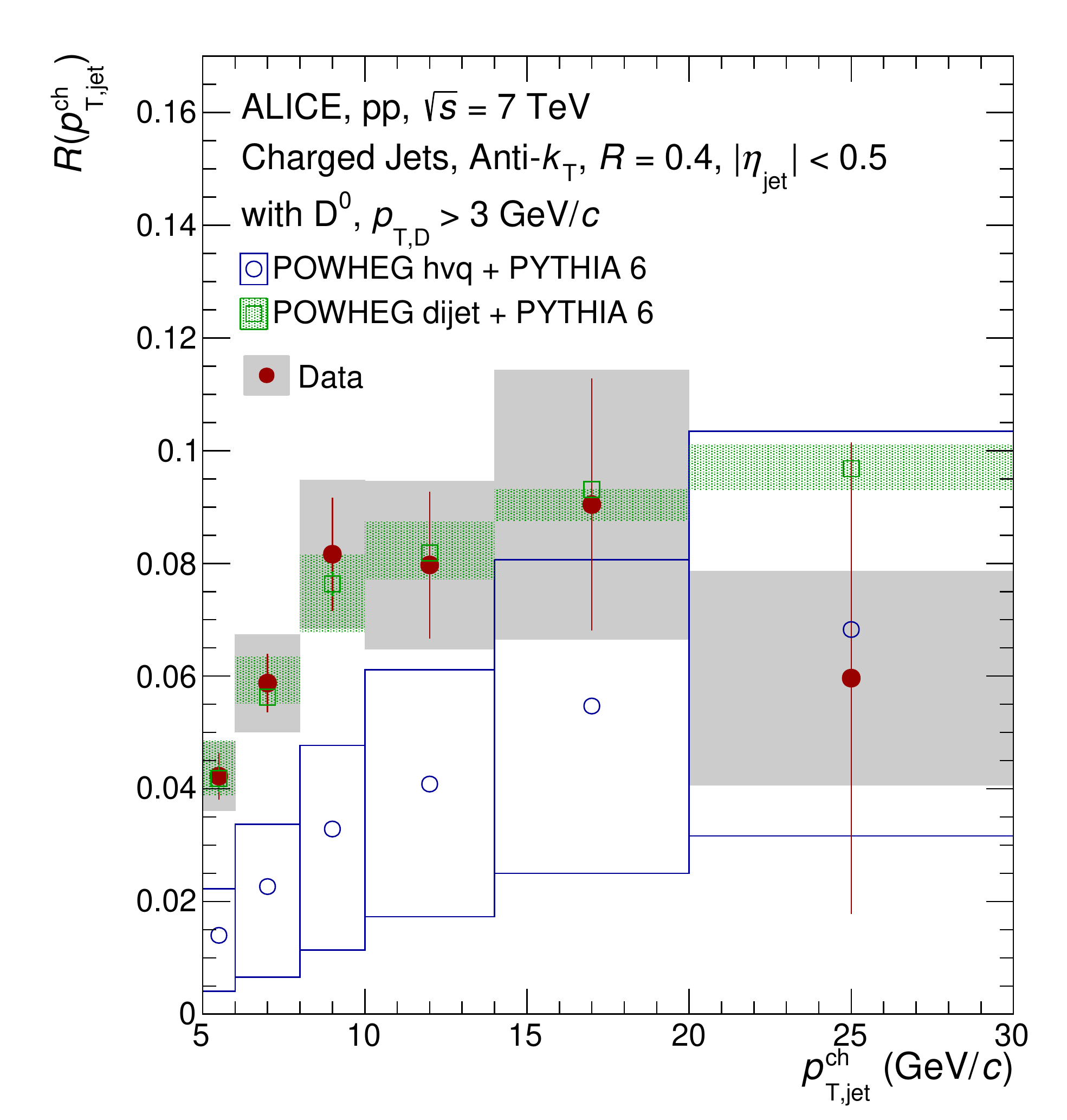}
\end{subfigure}
\caption{\pt-differential cross section of charm jets tagged with \Dzero\ mesons (left) and its ratio to the inclusive jet cross section (right) in \pp\ collisions at $\s=7$~TeV.
The solid red circles show the data with their systematic uncertainties represented by the grey boxes.
The measurements are compared with the \mbox{POWHEG} heavy-quark (open circles) and di-jet (open squares) implementations.}
\label{fig:cross_section_powheg}
\end{figure}

Figure~\ref{fig:cross_section_mc} (left) shows the \pt-differential cross section of charm jets containing a \Dzero\ meson in \pp\ collisions at $\s=7$~TeV.
The cross section is shown for $5<\ptchjet<30$~\GeVc.
The \Dzero\ mesons used to tag the jets have a minimum transverse momentum $\ptd>3$~\GeVc.
Figure~\ref{fig:cross_section_mc} (right) shows the rate of the \Dzero-meson tagged jets over the inclusive jet production as a function of \ptchjet:
\begin{equation}
R(\ptchjet)=\frac{N_{\rm \Dzero\,jet}(\ptchjet)}{N_{\rm jet}(\ptchjet)}.
\end{equation}
The inclusive jet production cross section in \pp\ collisions at $\s=7$~TeV was reported by ALICE in~\cite{ALICE:2015e} and more recently in~\cite{ALICE:2018b},
where the kinematic reach was extended down to $\ptchjet=5$~\GeVc.
The rate increases from about $0.04$ to about $0.08$ in the range $5<\ptchjet<10$~\GeVc; 
it then tends to flatten at a value around $0.08$ in the range $8<\ptchjet<20$~\GeVc.
According to Monte Carlo simulations based on POWHEG + PYTHIA~6, the increase of the charm-jet fraction in the interval $5<\pt<8~\GeVc$ is not due to the requirement of $\ptd>3$~\GeVc.

The measurements are compared with PYTHIA~6.4.28~\cite{Sjostrand:2014} (Perugia-2011 tune~\cite{Skands:2010a}), 
PYTHIA~8.2.1~\cite{Sjostrand:2014} (Monash-2013 tune~\cite{Skands:2014}) and Herwig 7~\cite{Bahr:2008, Bellm:2016} (\texttt{MEPP2QQ} and \texttt{MEMinBias}).
Both versions of PYTHIA overestimate the yield by a factor $\approx 1.5$ which appears to be approximately constant in the measured \ptchjet\ range
as shown in Fig.~\ref{fig:cross_section_mc} (left).
However, since they also overestimate the inclusive jet cross section by a similar amount~\cite{ALICE:2018b},
they provide a good description of the ratio of \Dzero-meson tagged jets over the inclusive jet production, as shown in Fig.~\ref{fig:cross_section_mc} (right).
For the purpose of this comparison the most prominent difference between the Herwig processes \texttt{MEPP2QQ} and \texttt{MEMinBias} is that
the former implements massive quarks in the matrix element calculations, whereas the latter treats all quarks as massless partons~\cite{Bahr:2008}.
When calculating the ratio, the inclusive jet cross section from the Herwig \texttt{MEMinBias} process was used for both the \texttt{MEMinBias} and the \texttt{MEPP2QQ} processes in the numerator.
Both Herwig processes tend to overestimate the measured cross section, with \texttt{MEPP2QQ} describing the data better.
The \texttt{MEMinBias} process reproduces well the ratio to the inclusive jet cross section.

The measurement is also compared with two NLO pQCD calculations obtained with the POWHEG-BOX V2 framework~\cite{Nason:2004a,Frixione:2007a, Alioli:2010a}, 
matched with PYTHIA~6 (Perugia-2011 tune)
for the generation of the parton shower and of the non-perturbative aspects of the simulation, such as hadronization of colored partons and generation of the underlying event.
The theoretical uncertainties were estimated by varying the renormalization and factorization scales ($0.5\mu_0\leq\mu_{\rm F, R}\leq2.0\mu_0$ with $0.5\leq\mu_{\rm R}/\mu_{\rm F}\leq2.0$), 
the mass of the charm quark ($m_{\rm c}=1.3,~1.7$~\GeVcsq\ with $m_{\rm c, 0}=1.5$~\GeVcsq) 
and the parton distribution function (central points: CT10nlo; variation: MSTW2008nlo68cl~\cite{Martin:2009}).
Two process implementations of the POWHEG framework were employed: the heavy-quark~\cite{Frixione:2007b} and the di-jet implementation~\cite{Alioli:2010b}.
As shown in Fig.~\ref{fig:cross_section_powheg} (left), good agreement is found within the theoretical and 
experimental uncertainties between the measured \pt-differential cross section and the cross section obtained with the POWHEG heavy-quark implementation.
The POWHEG di-jet implementation
systematically overestimates the production yield by a constant factor of $\approx 1.5$.
The right panel of Fig.~\ref{fig:cross_section_powheg} shows a comparison of the ratio of the \Dzero-meson tagged jet yield over the inclusive jet yield, for the data and POWHEG.
The inclusive jet yield was obtained with the POWHEG di-jet implementation for both presented POWHEG ratio cases.
In the calculation of the theoretical systematic uncertainties on the cross-section ratios, the variations of the perturbative scales and of the PDF were applied consistently in the numerator and in the denominator.
The measured ratio is found to be in agreement with the di-jet implementation, while the heavy-quark implementation systematically underestimates the ratio.
It is worth remarking that the excellent agreement between the data and the POWHEG di-jet implementation for $R(\ptchjet)$ means that
it overestimates both the \Dzero-meson tagged and the inclusive jet cross sections by a similar factor.

\begin{figure}[tbh]
\centering
\begin{subfigure}[b]{0.45\textwidth}
\includegraphics[width=\textwidth]{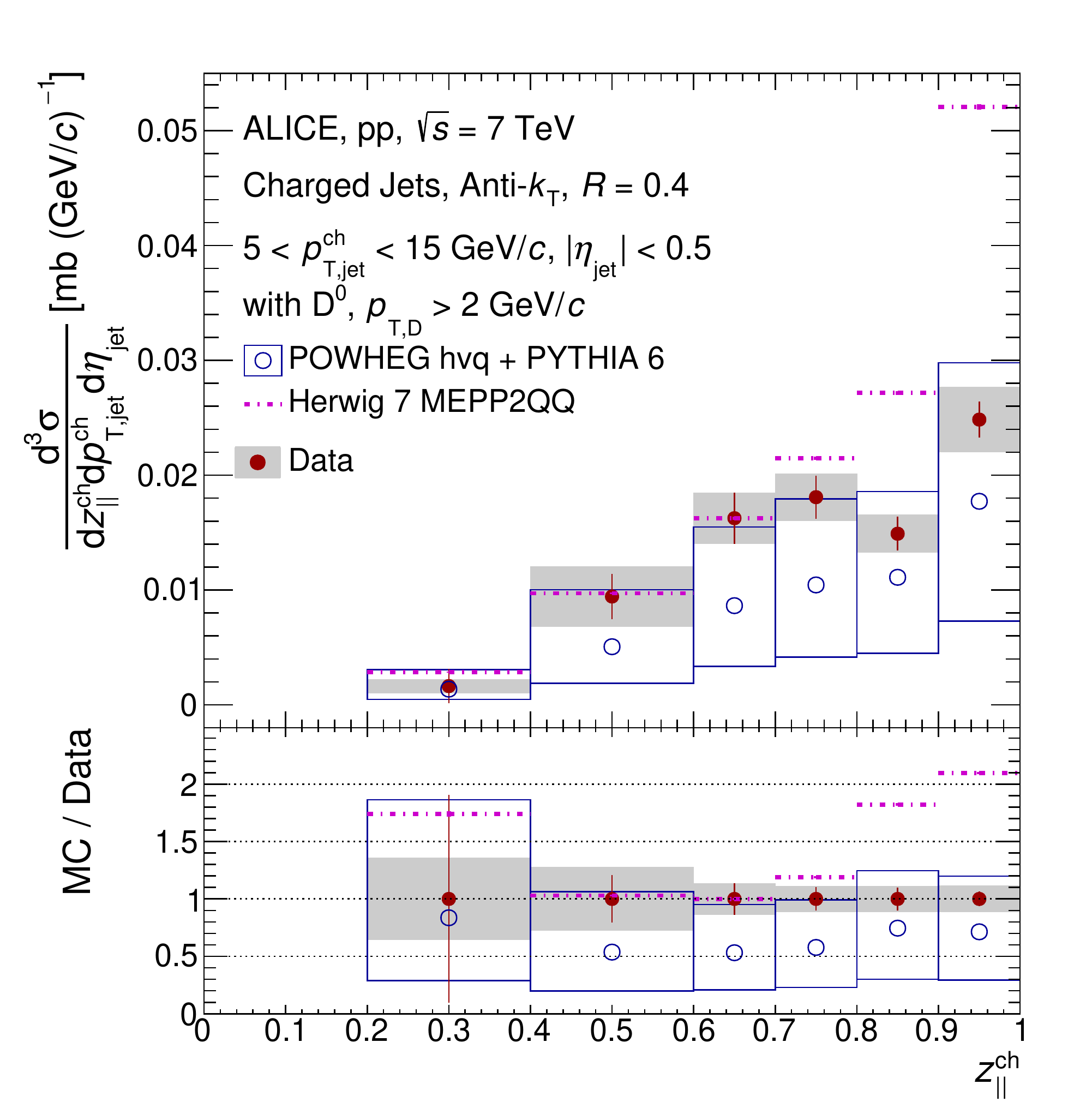}
\end{subfigure}\quad
\begin{subfigure}[b]{0.45\textwidth}
\includegraphics[width=\textwidth]{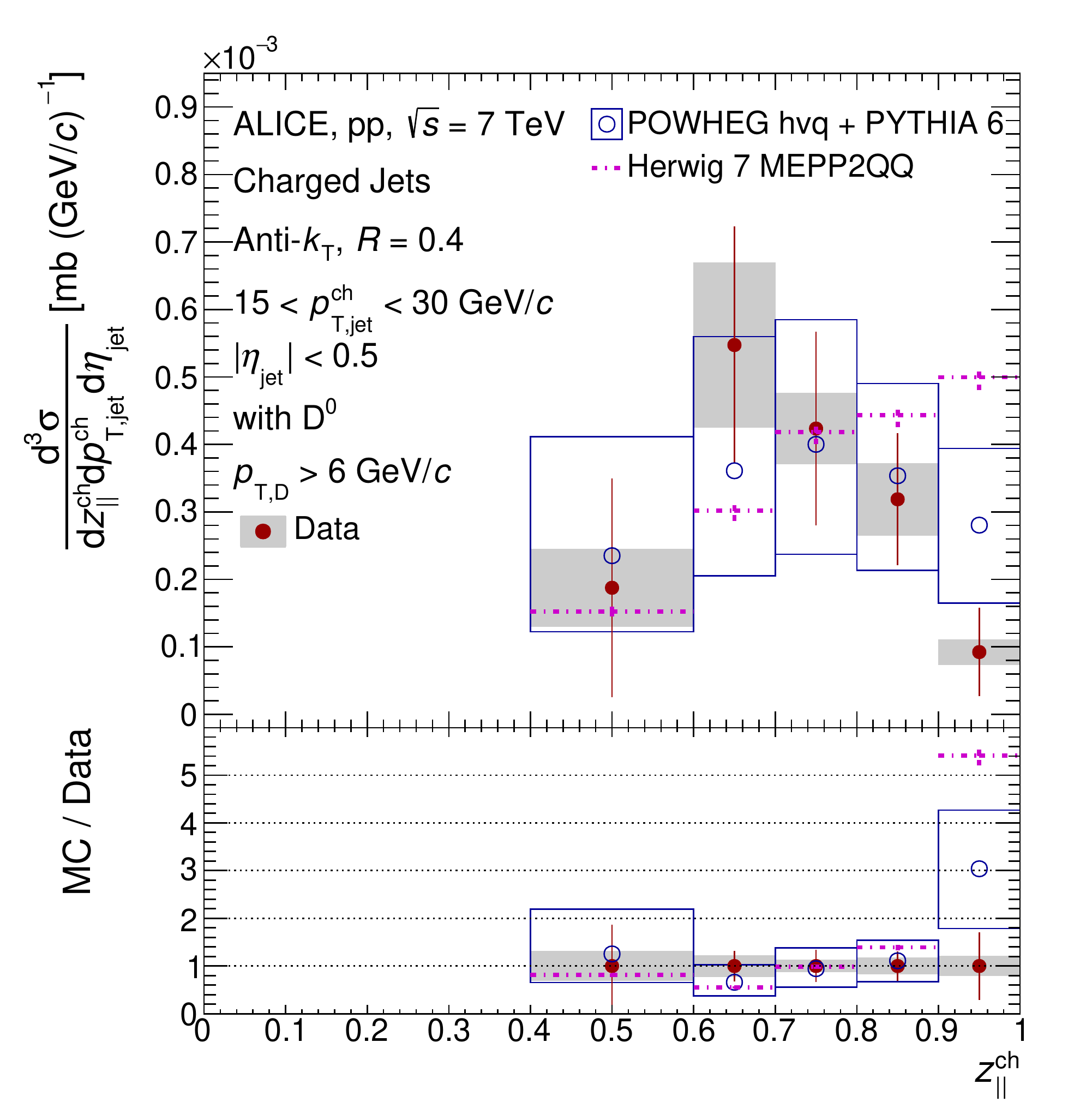}
\end{subfigure}
\caption{\zpar-differential cross section of \Dzero-meson tagged track-based jets
in \pp\ collisions at $\s=7$~TeV with $5<\ptchjet<15$~\GeVc\ (left) and $15<\ptchjet<30$~\GeVc\ (right).
The solid red circles show the data with their systematic uncertainties represented by gray boxes.
The measurements are compared with the \mbox{POWHEG} heavy-quark implementation (open circles) and
with Herwig 7 \texttt{MEPP2QQ} (dashed-dotted lines).}
\label{fig:cross_section_vs_z}
\end{figure}

\begin{figure}[tbh]
\centering
\begin{subfigure}[b]{0.45\textwidth}
\includegraphics[width=\textwidth]{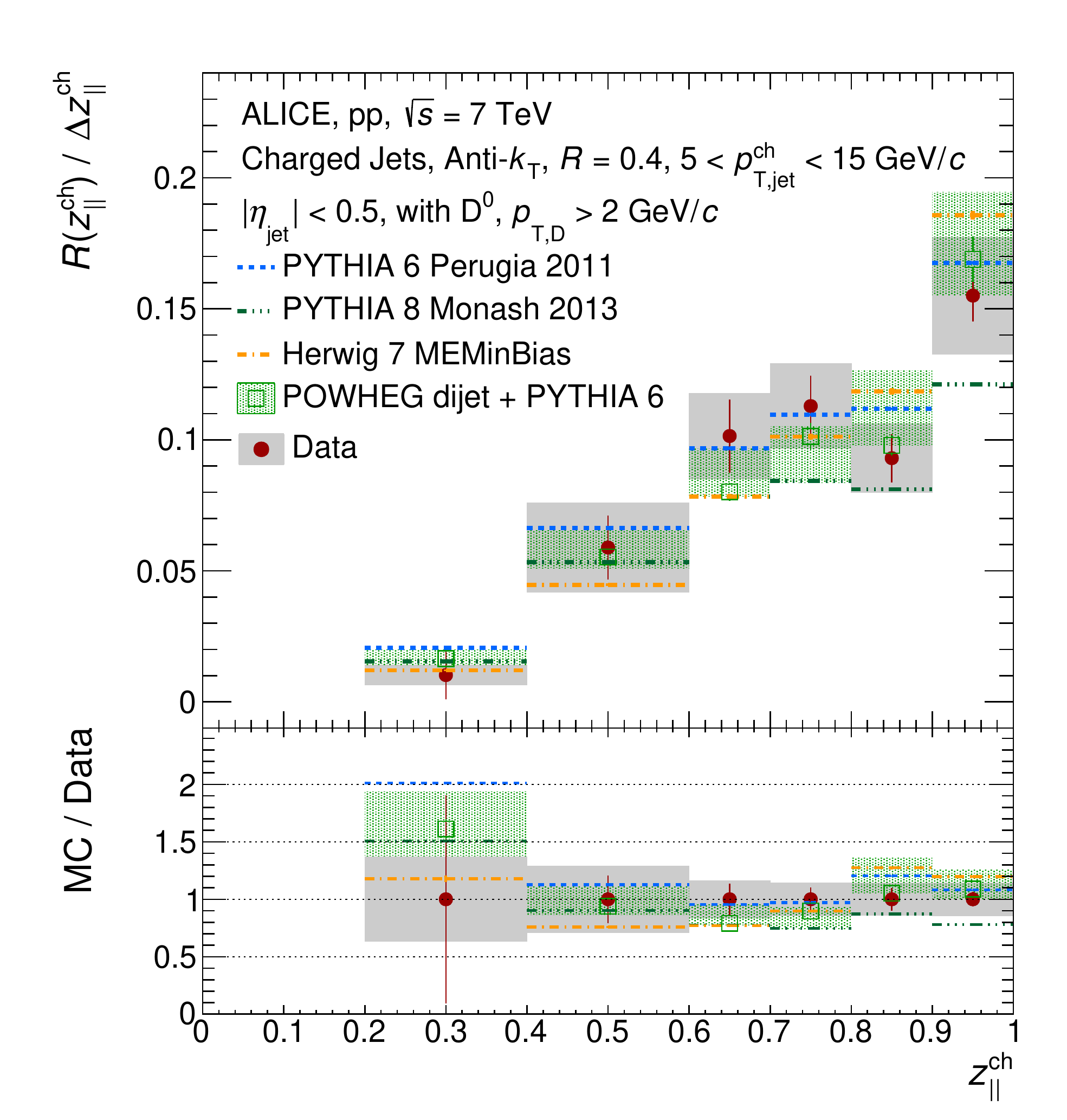}
\end{subfigure}\quad
\begin{subfigure}[b]{0.45\textwidth}
\includegraphics[width=\textwidth]{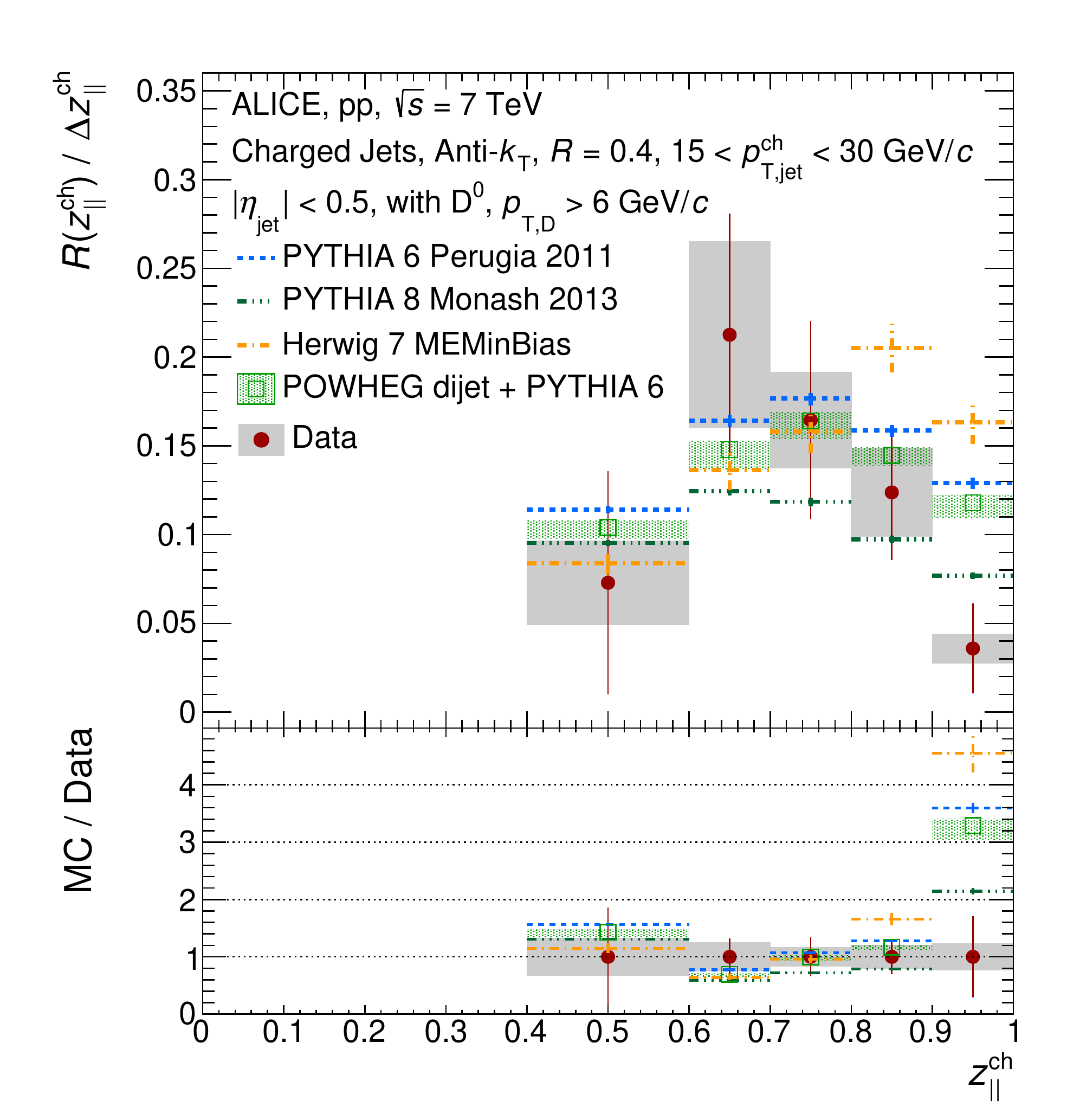}
\end{subfigure}
\caption{(Colours online) Rate of \Dzero-meson tagged track-based jets as a function of the jet momentum fraction carried by the \Dzero\ mesons in the direction of the jet axis
in \pp\ collisions at $\s=7$~TeV with $5<\ptchjet<15$~\GeVc\ (left) and $15<\ptchjet<30$~\GeVc\ (right).
The solid red circles show the data with their systematic uncertainties represented by gray boxes.
The measurements are compared with \mbox{PYTHIA~6} Perugia 2011 (blue), \mbox{PYTHIA~8} Monash 2013 (green dashed lines),
the \mbox{POWHEG} di-jet implementation (open squares) and Herwig 7 \texttt{MEQCDMinBias} (orange). }
\label{fig:rate_vs_z}
\end{figure}

Figure~\ref{fig:cross_section_vs_z} shows the \zpar-differential cross section of \Dzero-meson tagged jets
for $5<\ptchjet<15$~\GeVc\ (left) and for $15<\ptchjet<30$~\GeVc\ (right).
The \Dzero\ mesons used to tag the jets have a minimum transverse momentum $\ptd>2$~\GeVc\ for $5<\ptchjet<15$~\GeVc\ and $\ptd>6$~\GeVc\ for $15<\ptchjet<30$~\GeVc.
These kinematic selections allow one to fully access the \zpar\ distribution in $0.4<\zpar<1.0$ for both jet momentum intervals.
In the range $0.2<\zpar<0.4$, shown only for the lower jet momentum interval, the yield is biased by the missing contribution of \Dzero\ mesons with $1<\ptd<2$~\GeVc\footnote{The
bias was studied in the Monte Carlo simulations that are compared to the data and it was found to be smaller than the experimental uncertainties of the data.
The simulations used in the comparisons showed in this paper employs the same kinematic selections used in data.}
In the lower \ptchjet\ interval, a pronounced peak at $\zpar\approx1$ is observed.
This peak is populated by jets in which the \Dzero\ meson is the only constituent.
In the higher \ptchjet\ interval single-constituent jets are much rarer and the peak at $\zpar\approx1$ disappears.
In general, as \ptchjet\ increases the fragmentation becomes softer, a feature that has been observed also in inclusive jet measurements~\cite{ATLAS:2011a}.

In Fig.~\ref{fig:cross_section_vs_z}, the data are compared with simulations obtained with the POWHEG heavy-quark implementation and the Herwig 7 \texttt{MEPP2QQ} process,
both of which showed the best agreement with the \Dzero-meson tagged jet \pt-differential cross section in Figs.~\ref{fig:cross_section_mc} and \ref{fig:cross_section_powheg}.

The same data are shown in Fig.~\ref{fig:rate_vs_z} with a different normalization choice. 
The \zpar-differential cross section was divided by the inclusive jet cross section integrated in the corresponding \ptchjet\ interval:
\begin{equation}
\label{eq:mom_frac}
R(\ptchjet,\zpar)=\frac{N_{\rm \Dzero\,jet}(\ptchjet,\zpar)}{N_{\rm jet}(\ptchjet)}.
\end{equation}

In this case the data are compared with the POWHEG dijet implementation,
both versions of PYTHIA and the Herwig 7 \texttt{MEMinBias} process, which showed the best agreement with the ratio of the \Dzero-meson tagged jet cross section over the inclusive jet cross section
in Figs.~\ref{fig:cross_section_mc} and \ref{fig:cross_section_powheg}.
The choice of these two normalization approaches facilitate the comparison between data and simulation of the shapes of the \zpar\ distributions.

All models show an overall good agreement with the \zpar-differential data for jets with $5<\ptchjet<15$~\GeVc, with the only exception of Herwig 7 \texttt{MEPP2QQ},
which features a substantially harder fragmentation of \Dzero\ mesons in jets.
For jets with $15<\ptchjet<30$~\GeVc\, the models can describe the data quite well within large uncertainties.
A depletion is observed in the last \zpar\ bin in data compared to all models; however the discrepancy is only slightly larger than $1$~$\sigma$.

The measurement of the \zpar\ distribution for $15<\ptchjet<30$~\GeVc\ partially overlaps with the \Dstar\ in-jet fragmentation data reported by ATLAS in~\cite{ATLAS:2012d} for $25<\ptjet<30$~\GeVc.
While the jet measurement reported here includes only charged tracks,
the ATLAS measurement also includes neutral constituents of the jets,
and this difference should be taken into account while comparing our measurement with that of ATLAS.
The mean transverse momenta of the track-based jets considered in our analysis are $7.53 \pm 0.07$~\GeVc\ for $5<\ptchjet<15$~\GeVc\ and  $19.5 \pm 0.1$~\GeVc\ for $15<\ptchjet<30$~\GeVc.
Using a POWHEG + PYTHIA~6 simulation, it was estimated the transverse momentum of \Dzero-meson tagged jets increases on average by $12$\% for $5<\ptchjet<15$~\GeVc\ and by $14$\% for $15<\ptchjet<30$~\GeVc\ when neutral particles are included.
Furthermore, ATLAS reported jets reconstructed with a resolution parameter $R=0.6$ instead of $R=0.4$ used throughout this work.
Finally, in the case of the ATLAS measurement the contribution from the b-hadron feed-down was not subtracted.
\mbox{ATLAS} observed a large disagreement between data and various Monte Carlo event generators, including PYTHIA~6 and POWHEG di-jet.
Our data indicate a much better agreement with the simulations, however experimental uncertainties are large.

%% file: conclusions.tex
\section{Conclusions}
\label{sec:conclusions}
The measurement of charm jet production and fragmentation in \pp\ collisions at $\s=7$~TeV, in which charm jets are tagged using fully reconstructed \Dzero\ mesons, was presented in this paper.
The \Dzero-meson tagged jet \pt-differential cross section was reported in the range $5<\ptchjet<30$~\GeVc.
The fraction of charged jets containing a \mbox{\Dzero-meson}
increases with $\ptchjet$ from $0.042 \pm 0.004\, \mathrm{(stat)} \pm 0.006\, \mathrm{(syst)}$ to $0.080 \pm 0.009\, \mathrm{(stat)} \pm 0.008\, \mathrm{(syst)}$.
The cross section of \Dzero-meson tagged jets was reported also differentially as a function of the jet 
momentum fraction carried by the \Dzero\ meson in the direction of the jet axis (\zpar)
for two ranges of jet transverse momenta, $5<\ptchjet<15$~\GeVc\ and $15<\ptchjet<30$~\GeVc\ in the ranges $0.2<\zpar<1$ and $0.4<\zpar<1.0$, respectively.

The data were compared with PYTHIA~6, PYTHIA~8 and Herwig 7. Both versions of PYTHIA are able to describe reasonably well the ratio to inclusive jets, but not the cross section.
The Herwig 7 implementation of the heavy-quark production process can describe both the \pt-differential cross section of \Dzero-meson tagged jets and its ratio to the inclusive jet cross section.
The measurement was also compared with two NLO pQCD calculations obtained with the POWHEG heavy-quark and di-jet implementation.
The POWHEG heavy-quark implementation can reproduce the absolute cross section; when comparing with the ratio to inclusive jets (using the POWHEG di-jet implementation for the inclusive jets),
it significantly underestimates the data. When the POWHEG di-jet implementation is used both for the charm and the inclusive jet production the agreement to the ratio is restored.
All reported models can describe the measured \Dzero-meson tagged jet fragmentation within the uncertainties.
A small tension between the data and simulations is observed for $15<\ptchjet<30$~\GeVc, with the data favouring a slightly softer fragmentation.

The experimental uncertainties are dominated by the limited statistics: the analysis of larger data samples,
like those collected by ALICE in \pp\ collisions at $\s=5$ and 13~TeV, may allow for more differential measurements and a more conclusive comparison between data and theoretical expectations.
Although the uncertainties are still sizeable, agreement of the measurements with calculations provided by PYTHIA, Herwig and POWHEG indicates
that the observables studied in this work are well described by pQCD and they can therefore be exploited to address possible modifications
to the charm jet production and internal structure induced by the Quark--Gluon Plasma medium formed in heavy-ion collisions.

%% file: fa_2019-04-09.tex

The ALICE Collaboration would like to thank all its engineers and technicians for their invaluable contributions to the construction of the experiment and the CERN accelerator teams for the outstanding performance of the LHC complex.
The ALICE Collaboration gratefully acknowledges the resources and support provided by all Grid centres and the Worldwide LHC Computing Grid (WLCG) collaboration.
The ALICE Collaboration acknowledges the following funding agencies for their support in building and running the ALICE detector:
A. I. Alikhanyan National Science Laboratory (Yerevan Physics Institute) Foundation (ANSL), State Committee of Science and World Federation of Scientists (WFS), Armenia;
Austrian Academy of Sciences, Austrian Science Fund (FWF): [M 2467-N36] and Nationalstiftung f\"{u}r Forschung, Technologie und Entwicklung, Austria;
Ministry of Communications and High Technologies, National Nuclear Research Center, Azerbaijan;
Conselho Nacional de Desenvolvimento Cient\'{\i}fico e Tecnol\'{o}gico (CNPq), Universidade Federal do Rio Grande do Sul (UFRGS), Financiadora de Estudos e Projetos (Finep) and Funda\c{c}\~{a}o de Amparo \`{a} Pesquisa do Estado de S\~{a}o Paulo (FAPESP), Brazil;
Ministry of Science \& Technology of China (MSTC), National Natural Science Foundation of China (NSFC) and Ministry of Education of China (MOEC) , China;
Croatian Science Foundation and Ministry of Science and Education, Croatia;
Centro de Aplicaciones Tecnol\'{o}gicas y Desarrollo Nuclear (CEADEN), Cubaenerg\'{\i}a, Cuba;
Ministry of Education, Youth and Sports of the Czech Republic, Czech Republic;
The Danish Council for Independent Research | Natural Sciences, the Carlsberg Foundation and Danish National Research Foundation (DNRF), Denmark;
Helsinki Institute of Physics (HIP), Finland;
Commissariat \`{a} l'Energie Atomique (CEA), Institut National de Physique Nucl\'{e}aire et de Physique des Particules (IN2P3) and Centre National de la Recherche Scientifique (CNRS) and R\'{e}gion des  Pays de la Loire, France;
Bundesministerium f\"{u}r Bildung und Forschung (BMBF) and GSI Helmholtzzentrum f\"{u}r Schwerionenforschung GmbH, Germany;
General Secretariat for Research and Technology, Ministry of Education, Research and Religions, Greece;
National Research, Development and Innovation Office, Hungary;
Department of Atomic Energy Government of India (DAE), Department of Science and Technology, Government of India (DST), University Grants Commission, Government of India (UGC) and Council of Scientific and Industrial Research (CSIR), India;
Indonesian Institute of Science, Indonesia;
Centro Fermi - Museo Storico della Fisica e Centro Studi e Ricerche Enrico Fermi and Istituto Nazionale di Fisica Nucleare (INFN), Italy;
Institute for Innovative Science and Technology , Nagasaki Institute of Applied Science (IIST), Japan Society for the Promotion of Science (JSPS) KAKENHI and Japanese Ministry of Education, Culture, Sports, Science and Technology (MEXT), Japan;
Consejo Nacional de Ciencia (CONACYT) y Tecnolog\'{i}a, through Fondo de Cooperaci\'{o}n Internacional en Ciencia y Tecnolog\'{i}a (FONCICYT) and Direcci\'{o}n General de Asuntos del Personal Academico (DGAPA), Mexico;
Nederlandse Organisatie voor Wetenschappelijk Onderzoek (NWO), Netherlands;
The Research Council of Norway, Norway;
Commission on Science and Technology for Sustainable Development in the South (COMSATS), Pakistan;
Pontificia Universidad Cat\'{o}lica del Per\'{u}, Peru;
Ministry of Science and Higher Education and National Science Centre, Poland;
Korea Institute of Science and Technology Information and National Research Foundation of Korea (NRF), Republic of Korea;
Ministry of Education and Scientific Research, Institute of Atomic Physics and Ministry of Research and Innovation and Institute of Atomic Physics, Romania;
Joint Institute for Nuclear Research (JINR), Ministry of Education and Science of the Russian Federation, National Research Centre Kurchatov Institute, Russian Science Foundation and Russian Foundation for Basic Research, Russia;
Ministry of Education, Science, Research and Sport of the Slovak Republic, Slovakia;
National Research Foundation of South Africa, South Africa;
Swedish Research Council (VR) and Knut \& Alice Wallenberg Foundation (KAW), Sweden;
European Organization for Nuclear Research, Switzerland;
National Science and Technology Development Agency (NSDTA), Suranaree University of Technology (SUT) and Office of the Higher Education Commission under NRU project of Thailand, Thailand;
Turkish Atomic Energy Agency (TAEK), Turkey;
National Academy of  Sciences of Ukraine, Ukraine;
Science and Technology Facilities Council (STFC), United Kingdom;
National Science Foundation of the United States of America (NSF) and United States Department of Energy, Office of Nuclear Physics (DOE NP), United States of America.

%% file: 2019-04-09-Alice_Authorlist_2019-Apr-09.tex

\begingroup
\small
\begin{flushleft}
S.~Acharya\Irefn{org141}\And 
D.~Adamov\'{a}\Irefn{org93}\And 
S.P.~Adhya\Irefn{org141}\And 
A.~Adler\Irefn{org74}\And 
J.~Adolfsson\Irefn{org80}\And 
M.M.~Aggarwal\Irefn{org98}\And 
G.~Aglieri Rinella\Irefn{org34}\And 
M.~Agnello\Irefn{org31}\And 
N.~Agrawal\Irefn{org10}\And 
Z.~Ahammed\Irefn{org141}
% ...
\And
\renewcommand\labelenumi{\textsuperscript{\theenumi}~}

\section*{Affiliation notes}
\renewcommand\theenumi{\roman{enumi}}
\begin{Authlist}
\item \Adef{org*}Deceased
\item \Adef{orgI}Dipartimento DET del Politecnico di Torino, Turin, Italy
\item \Adef{orgII}M.V. Lomonosov Moscow State University, D.V. Skobeltsyn Institute of Nuclear, Physics, Moscow, Russia
\item \Adef{orgIII}Department of Applied Physics, Aligarh Muslim University, Aligarh, India
\item \Adef{orgIV}Institute of Theoretical Physics, University of Wroclaw, Poland
\end{Authlist}

\section*{Collaboration Institutes}
\renewcommand\theenumi{\arabic{enumi}~}
\begin{Authlist}
\item \Idef{org1}A.I. Alikhanyan National Science Laboratory (Yerevan Physics Institute) Foundation, Yerevan, Armenia
\item \Idef{org2}Bogolyubov Institute for Theoretical Physics, National Academy of Sciences of Ukraine, Kiev, Ukraine
\item \Idef{org3}Bose Institute, Department of Physics  and Centre for Astroparticle Physics and Space Science (CAPSS), Kolkata, India
\item \Idef{org4}Budker Institute for Nuclear Physics, Novosibirsk, Russia
\item \Idef{org5}California Polytechnic State University, San Luis Obispo, California, United States
\item \Idef{org6}Central China Normal University, Wuhan, China
\item \Idef{org7}Centre de Calcul de l'IN2P3, Villeurbanne, Lyon, France
\item \Idef{org8}Centro de Aplicaciones Tecnol\'{o}gicas y Desarrollo Nuclear (CEADEN), Havana, Cuba
\item \Idef{org9}Centro de Investigaci\'{o}n y de Estudios Avanzados (CINVESTAV), Mexico City and M\'{e}rida, Mexico
\item \Idef{org10}Centro Fermi - Museo Storico della Fisica e Centro Studi e Ricerche ``Enrico Fermi', Rome, Italy
\item \Idef{org11}Chicago State University, Chicago, Illinois, United States
\item \Idef{org12}China Institute of Atomic Energy, Beijing, China
\item \Idef{org13}Chonbuk National University, Jeonju, Republic of Korea
\item \Idef{org14}Comenius University Bratislava, Faculty of Mathematics, Physics and Informatics, Bratislava, Slovakia
\item \Idef{org15}COMSATS University Islamabad, Islamabad, Pakistan
\item \Idef{org16}Creighton University, Omaha, Nebraska, United States
\item \Idef{org17}Department of Physics, Aligarh Muslim University, Aligarh, India
\item \Idef{org18}Department of Physics, Pusan National University, Pusan, Republic of Korea
\item \Idef{org19}Department of Physics, Sejong University, Seoul, Republic of Korea
\item \Idef{org20}Department of Physics, University of California, Berkeley, California, United States
\item \Idef{org21}Department of Physics, University of Oslo, Oslo, Norway
\item \Idef{org22}Department of Physics and Technology, University of Bergen, Bergen, Norway
\item \Idef{org23}Dipartimento di Fisica dell'Universit\`{a} 'La Sapienza' and Sezione INFN, Rome, Italy
\item \Idef{org24}Dipartimento di Fisica dell'Universit\`{a} and Sezione INFN, Cagliari, Italy
\item \Idef{org25}Dipartimento di Fisica dell'Universit\`{a} and Sezione INFN, Trieste, Italy
\item \Idef{org26}Dipartimento di Fisica dell'Universit\`{a} and Sezione INFN, Turin, Italy
\item \Idef{org27}Dipartimento di Fisica e Astronomia dell'Universit\`{a} and Sezione INFN, Bologna, Italy
\item \Idef{org28}Dipartimento di Fisica e Astronomia dell'Universit\`{a} and Sezione INFN, Catania, Italy
\item \Idef{org29}Dipartimento di Fisica e Astronomia dell'Universit\`{a} and Sezione INFN, Padova, Italy
\item \Idef{org30}Dipartimento di Fisica `E.R.~Caianiello' dell'Universit\`{a} and Gruppo Collegato INFN, Salerno, Italy
\item \Idef{org31}Dipartimento DISAT del Politecnico and Sezione INFN, Turin, Italy
\item \Idef{org32}Dipartimento di Scienze e Innovazione Tecnologica dell'Universit\`{a} del Piemonte Orientale and INFN Sezione di Torino, Alessandria, Italy
\item \Idef{org33}Dipartimento Interateneo di Fisica `M.~Merlin' and Sezione INFN, Bari, Italy
\item \Idef{org34}European Organization for Nuclear Research (CERN), Geneva, Switzerland
\item \Idef{org35}Faculty of Electrical Engineering, Mechanical Engineering and Naval Architecture, University of Split, Split, Croatia
\item \Idef{org36}Faculty of Engineering and Science, Western Norway University of Applied Sciences, Bergen, Norway
\item \Idef{org37}Faculty of Nuclear Sciences and Physical Engineering, Czech Technical University in Prague, Prague, Czech Republic
\item \Idef{org38}Faculty of Science, P.J.~\v{S}af\'{a}rik University, Ko\v{s}ice, Slovakia
\item \Idef{org39}Frankfurt Institute for Advanced Studies, Johann Wolfgang Goethe-Universit\"{a}t Frankfurt, Frankfurt, Germany
\item \Idef{org40}Gangneung-Wonju National University, Gangneung, Republic of Korea
\item \Idef{org41}Gauhati University, Department of Physics, Guwahati, India
\item \Idef{org42}Helmholtz-Institut f\"{u}r Strahlen- und Kernphysik, Rheinische Friedrich-Wilhelms-Universit\"{a}t Bonn, Bonn, Germany
\item \Idef{org43}Helsinki Institute of Physics (HIP), Helsinki, Finland
\item \Idef{org44}High Energy Physics Group,  Universidad Aut\'{o}noma de Puebla, Puebla, Mexico
\item \Idef{org45}Hiroshima University, Hiroshima, Japan
\item \Idef{org46}Hochschule Worms, Zentrum  f\"{u}r Technologietransfer und Telekommunikation (ZTT), Worms, Germany
\item \Idef{org47}Horia Hulubei National Institute of Physics and Nuclear Engineering, Bucharest, Romania
\item \Idef{org48}Indian Institute of Technology Bombay (IIT), Mumbai, India
\item \Idef{org49}Indian Institute of Technology Indore, Indore, India
\item \Idef{org50}Indonesian Institute of Sciences, Jakarta, Indonesia
\item \Idef{org51}INFN, Laboratori Nazionali di Frascati, Frascati, Italy
\item \Idef{org52}INFN, Sezione di Bari, Bari, Italy
\item \Idef{org53}INFN, Sezione di Bologna, Bologna, Italy
\item \Idef{org54}INFN, Sezione di Cagliari, Cagliari, Italy
\item \Idef{org55}INFN, Sezione di Catania, Catania, Italy
\item \Idef{org56}INFN, Sezione di Padova, Padova, Italy
\item \Idef{org57}INFN, Sezione di Roma, Rome, Italy
\item \Idef{org58}INFN, Sezione di Torino, Turin, Italy
\item \Idef{org59}INFN, Sezione di Trieste, Trieste, Italy
\item \Idef{org60}Inha University, Incheon, Republic of Korea
\item \Idef{org61}Institut de Physique Nucl\'{e}aire d'Orsay (IPNO), Institut National de Physique Nucl\'{e}aire et de Physique des Particules (IN2P3/CNRS), Universit\'{e} de Paris-Sud, Universit\'{e} Paris-Saclay, Orsay, France
\item \Idef{org62}Institute for Nuclear Research, Academy of Sciences, Moscow, Russia
\item \Idef{org63}Institute for Subatomic Physics, Utrecht University/Nikhef, Utrecht, Netherlands
\item \Idef{org64}Institute for Theoretical and Experimental Physics, Moscow, Russia
\item \Idef{org65}Institute of Experimental Physics, Slovak Academy of Sciences, Ko\v{s}ice, Slovakia
\item \Idef{org66}Institute of Physics, Homi Bhabha National Institute, Bhubaneswar, India
\item \Idef{org67}Institute of Physics of the Czech Academy of Sciences, Prague, Czech Republic
\item \Idef{org68}Institute of Space Science (ISS), Bucharest, Romania
\item \Idef{org69}Institut f\"{u}r Kernphysik, Johann Wolfgang Goethe-Universit\"{a}t Frankfurt, Frankfurt, Germany
\item \Idef{org70}Instituto de Ciencias Nucleares, Universidad Nacional Aut\'{o}noma de M\'{e}xico, Mexico City, Mexico
\item \Idef{org71}Instituto de F\'{i}sica, Universidade Federal do Rio Grande do Sul (UFRGS), Porto Alegre, Brazil
\item \Idef{org72}Instituto de F\'{\i}sica, Universidad Nacional Aut\'{o}noma de M\'{e}xico, Mexico City, Mexico
\item \Idef{org73}iThemba LABS, National Research Foundation, Somerset West, South Africa
\item \Idef{org74}Johann-Wolfgang-Goethe Universit\"{a}t Frankfurt Institut f\"{u}r Informatik, Fachbereich Informatik und Mathematik, Frankfurt, Germany
\item \Idef{org75}Joint Institute for Nuclear Research (JINR), Dubna, Russia
\item \Idef{org76}Korea Institute of Science and Technology Information, Daejeon, Republic of Korea
\item \Idef{org77}KTO Karatay University, Konya, Turkey
\item \Idef{org78}Laboratoire de Physique Subatomique et de Cosmologie, Universit\'{e} Grenoble-Alpes, CNRS-IN2P3, Grenoble, France
\item \Idef{org79}Lawrence Berkeley National Laboratory, Berkeley, California, United States
\item \Idef{org80}Lund University Department of Physics, Division of Particle Physics, Lund, Sweden
\item \Idef{org81}Nagasaki Institute of Applied Science, Nagasaki, Japan
\item \Idef{org82}Nara Women{'}s University (NWU), Nara, Japan
\item \Idef{org83}National and Kapodistrian University of Athens, School of Science, Department of Physics , Athens, Greece
\item \Idef{org84}National Centre for Nuclear Research, Warsaw, Poland
\item \Idef{org85}National Institute of Science Education and Research, Homi Bhabha National Institute, Jatni, India
\item \Idef{org86}National Nuclear Research Center, Baku, Azerbaijan
\item \Idef{org87}National Research Centre Kurchatov Institute, Moscow, Russia
\item \Idef{org88}Niels Bohr Institute, University of Copenhagen, Copenhagen, Denmark
\item \Idef{org89}Nikhef, National institute for subatomic physics, Amsterdam, Netherlands
\item \Idef{org90}NRC Kurchatov Institute IHEP, Protvino, Russia
\item \Idef{org91}NRNU Moscow Engineering Physics Institute, Moscow, Russia
\item \Idef{org92}Nuclear Physics Group, STFC Daresbury Laboratory, Daresbury, United Kingdom
\item \Idef{org93}Nuclear Physics Institute of the Czech Academy of Sciences, \v{R}e\v{z} u Prahy, Czech Republic
\item \Idef{org94}Oak Ridge National Laboratory, Oak Ridge, Tennessee, United States
\item \Idef{org95}Ohio State University, Columbus, Ohio, United States
\item \Idef{org96}Petersburg Nuclear Physics Institute, Gatchina, Russia
\item \Idef{org97}Physics department, Faculty of science, University of Zagreb, Zagreb, Croatia
\item \Idef{org98}Physics Department, Panjab University, Chandigarh, India
\item \Idef{org99}Physics Department, University of Jammu, Jammu, India
\item \Idef{org100}Physics Department, University of Rajasthan, Jaipur, India
\item \Idef{org101}Physikalisches Institut, Eberhard-Karls-Universit\"{a}t T\"{u}bingen, T\"{u}bingen, Germany
\item \Idef{org102}Physikalisches Institut, Ruprecht-Karls-Universit\"{a}t Heidelberg, Heidelberg, Germany
\item \Idef{org103}Physik Department, Technische Universit\"{a}t M\"{u}nchen, Munich, Germany
\item \Idef{org104}Politecnico di Bari, Bari, Italy
\item \Idef{org105}Research Division and ExtreMe Matter Institute EMMI, GSI Helmholtzzentrum f\"ur Schwerionenforschung GmbH, Darmstadt, Germany
\item \Idef{org106}Rudjer Bo\v{s}kovi\'{c} Institute, Zagreb, Croatia
\item \Idef{org107}Russian Federal Nuclear Center (VNIIEF), Sarov, Russia
\item \Idef{org108}Saha Institute of Nuclear Physics, Homi Bhabha National Institute, Kolkata, India
\item \Idef{org109}School of Physics and Astronomy, University of Birmingham, Birmingham, United Kingdom
\item \Idef{org110}Secci\'{o}n F\'{\i}sica, Departamento de Ciencias, Pontificia Universidad Cat\'{o}lica del Per\'{u}, Lima, Peru
\item \Idef{org111}Shanghai Institute of Applied Physics, Shanghai, China
\item \Idef{org112}St. Petersburg State University, St. Petersburg, Russia
\item \Idef{org113}Stefan Meyer Institut f\"{u}r Subatomare Physik (SMI), Vienna, Austria
\item \Idef{org114}SUBATECH, IMT Atlantique, Universit\'{e} de Nantes, CNRS-IN2P3, Nantes, France
\item \Idef{org115}Suranaree University of Technology, Nakhon Ratchasima, Thailand
\item \Idef{org116}Technical University of Ko\v{s}ice, Ko\v{s}ice, Slovakia
\item \Idef{org117}Technische Universit\"{a}t M\"{u}nchen, Excellence Cluster 'Universe', Munich, Germany
\item \Idef{org118}The Henryk Niewodniczanski Institute of Nuclear Physics, Polish Academy of Sciences, Cracow, Poland
\item \Idef{org119}The University of Texas at Austin, Austin, Texas, United States
\item \Idef{org120}Universidad Aut\'{o}noma de Sinaloa, Culiac\'{a}n, Mexico
\item \Idef{org121}Universidade de S\~{a}o Paulo (USP), S\~{a}o Paulo, Brazil
\item \Idef{org122}Universidade Estadual de Campinas (UNICAMP), Campinas, Brazil
\item \Idef{org123}Universidade Federal do ABC, Santo Andre, Brazil
\item \Idef{org124}University College of Southeast Norway, Tonsberg, Norway
\item \Idef{org125}University of Cape Town, Cape Town, South Africa
\item \Idef{org126}University of Houston, Houston, Texas, United States
\item \Idef{org127}University of Jyv\"{a}skyl\"{a}, Jyv\"{a}skyl\"{a}, Finland
\item \Idef{org128}University of Liverpool, Liverpool, United Kingdom
\item \Idef{org129}University of Science and Techonology of China, Hefei, China
\item \Idef{org130}University of Tennessee, Knoxville, Tennessee, United States
\item \Idef{org131}University of the Witwatersrand, Johannesburg, South Africa
\item \Idef{org132}University of Tokyo, Tokyo, Japan
\item \Idef{org133}University of Tsukuba, Tsukuba, Japan
\item \Idef{org134}Universit\'{e} Clermont Auvergne, CNRS/IN2P3, LPC, Clermont-Ferrand, France
\item \Idef{org135}Universit\'{e} de Lyon, Universit\'{e} Lyon 1, CNRS/IN2P3, IPN-Lyon, Villeurbanne, Lyon, France
\item \Idef{org136}Universit\'{e} de Strasbourg, CNRS, IPHC UMR 7178, F-67000 Strasbourg, France, Strasbourg, France
\item \Idef{org137}Universit\'{e} Paris-Saclay Centre d'Etudes de Saclay (CEA), IRFU, D\'{e}partment de Physique Nucl\'{e}aire (DPhN), Saclay, France
\item \Idef{org138}Universit\`{a} degli Studi di Foggia, Foggia, Italy
\item \Idef{org139}Universit\`{a} degli Studi di Pavia, Pavia, Italy
\item \Idef{org140}Universit\`{a} di Brescia, Brescia, Italy
\item \Idef{org141}Variable Energy Cyclotron Centre, Homi Bhabha National Institute, Kolkata, India
\item \Idef{org142}Warsaw University of Technology, Warsaw, Poland
\item \Idef{org143}Wayne State University, Detroit, Michigan, United States
\item \Idef{org144}Westf\"{a}lische Wilhelms-Universit\"{a}t M\"{u}nster, Institut f\"{u}r Kernphysik, M\"{u}nster, Germany
\item \Idef{org145}Wigner Research Centre for Physics, Hungarian Academy of Sciences, Budapest, Hungary
\item \Idef{org146}Yale University, New Haven, Connecticut, United States
\item \Idef{org147}Yonsei University, Seoul, Republic of Korea
\end{Authlist}
\endgroup